\documentclass{article}

\usepackage{PRIMEarxiv}

\usepackage[utf8]{inputenc} % allow utf-8 input
\usepackage[T1]{fontenc}    % use 8-bit T1 fonts
\usepackage{hyperref}       % hyperlinks
\usepackage{url}            % simple URL typesetting
\usepackage{booktabs}       % professional-quality tables
\usepackage{amsfonts}       % blackboard math symbols
\usepackage{nicefrac}       % compact symbols for 1/2, etc.
\usepackage{microtype}      % microtypography
\usepackage{lipsum}
\usepackage{fancyhdr}       % header
\usepackage{graphicx}       % graphics
\usepackage{amssymb}

\graphicspath{{media/}}     % organize your images and other figures under media/ folder

\AtBeginDocument{%
  \providecommand\BibTeX{{%
    \normalfont B\kern-0.5em{\scshape i\kern-0.25em b}\kern-0.8em\TeX}}}

\usepackage{comment}
\usepackage{algpseudocode,algorithm}
\usepackage{bm}
\usepackage{mathtools}
\algnewcommand{\TRUE}{\textbf{true}}
\algnewcommand{\FALSE}{\textbf{false}}
\newcommand{\N}{\mathbb{N}}
\newcommand{\R}{\mathbb{R}}
\newcommand{\PP}{\mathbb{P}}
\newcommand{\E}{\mathbb{E}}
\newcommand*\diff{\mathop{}\!\mathrm{d}}

\usepackage{color}
\usepackage{graphicx}

\usepackage{amsmath,amsthm,mathrsfs,amsfonts,mathtools,dsfont,bbm,anyfontsize}
\DeclareMathOperator*{\argmax}{arg\,max}

\newcommand{\equaref}[1]{(\ref{eq:#1})}

\newcommand{\sidebyside}[4]{
\begin{figure}[phtb]
\begin{minipage}[phtb]{7cm}
\includegraphics[width=0.98\textwidth]{#1.eps}
\caption{#2\label{fig:#1}}
\end{minipage}
\hfill \hspace{0.1cm}
\begin{minipage}[phtb]{7cm}
\includegraphics[width=0.98\textwidth]{#3.eps}
\caption{#4\label{fig:#3}}
\end{minipage}
\hfill
\vspace{-0mm}
\end{figure}
}

\newtheorem{definition}{Definition}[section]

\newtheorem{assumption}{Assumption}[section]

\newtheorem{lemma}{Lemma}[section]

\newtheorem{corollary}{Corollary}[section]

\newtheorem{proposition}{Proposition}[section]

\newtheorem{remark}{Remark}[section]
\newtheorem{example}{Example}[section]

\title{Reconciling the Quality vs Popularity Dichotomy in Online Cultural Markets
\thanks{\textit{\underline{Citation}}: 
\textbf{R. Gaeta, M. Garetto, G. Ruffo, and A. Flammini. 2022. Reconciling the Quality vs Popularity Dichotomy in Online Cultural Markets. ACM Trans. Inf. Syst. April 2022. DOI:https://doi.org/10.1145/3530790}} 
}

\author{
  Rossano Gaeta, Michele Garetto, Giancarlo Ruffo \\
  Dipartimento di Informatica \\
  Università degli Studi di Torino \\
  \texttt{\{rossano.gaeta, michele.garetto, giancarlo.ruffo\}@unito.it} \\
   \And
  Alessandro Flammini \\
  Network Science Institute \\
  Indiana University \\
  Bloomington, Indiana (US)\\
  \texttt{aflammin@indiana.edu} \\
}

\begin{document}
\maketitle

\begin{abstract}
  We propose a simple model of an idealized online cultural market in which $N$ items,
  endowed with a hidden quality metric, are recommended to users
  by a ranking algorithm possibly biased by the current items' popularity.
  Our goal is to better understand the underlying mechanisms of the well-known fact that popularity bias
  can prevent higher-quality items from becoming more popular than lower-quality items,
  producing an undesirable misalignment between quality and popularity rankings.
  We do so under the assumption that users, having limited time/attention, are able to
  discriminate the best-quality only within a random subset of the items.
  We discover the existence of a harmful regime in which improper use of popularity
  can seriously compromise the emergence of quality, and a benign regime in which
  wise use of popularity, coupled with a small discrimination effort on behalf of users,
  guarantees the perfect alignment of quality and popularity ranking. Our findings clarify the
  effects of algorithmic popularity bias on quality outcomes, and may inform the design of
  more principled mechanisms for techno-social cultural markets.
\end{abstract}


\section{Introduction}

It is a common experience to receive recommendations from an expert system.
It happens when we buy a book or music from Amazon, when we scroll the lists that Google returns in response to our queries, when we are shown potential friends on a social media site such as Twitter or Facebook, or when we search a generic news article on a news aggregation service.
Indeed, the availability of cultural and informational products, and the number of people we could potential connect to via social media is so vast that without recommender
systems~\cite{ricci_recommender_2010} we would have little chances to find what we like or need.
Such systems, broadly speaking, try to infer how pleased a user would be with a particular item, given her history of items consumption and taking into consideration the choices made from other, possibly similar, users, and return a list of personalized top-N items~\cite{10.1145/963770.963776}.
While there is a strong incentive to create systems that produce personalized recommendations and therefore pay special attention to choices made by users similar to the target of the recommendation, the popularity of an item (measured e.g. as the number of copies sold, views, downloads, likes) is a signal that is often leveraged to produce recommendations.
In \cite{jannach15}, for example, authors compare several recommendations strategies from different perspectives, including accuracy, catalog coverage and their bias to recommend popular items, and show that recent algorithmic techniques end up recommending mostly top sellers.

The notion that popularity is an indicator of quality is predicated on the notion of wisdom of the crowds~\cite{suroviecki_wisdom_2005},
the fact that an assessment by many independent - even non expert - individuals could be more precise/valuable/correct than that of few experts. Another justification for using popularity as proxy for quality
is that quality is a concept that is hard to define and measure, as it is intimately tied
to highly subjective notions such as beauty, novelty, and virality.

There is, however, ample empirical evidence showing that
social influence (resulting for example by popularity-based ranking algorithms)
can bias the success of different items in ways that do not reflect
their intrinsic quality \cite{lorenz}.
Intuitively, {\em popularity bias} can reinforce initial random fluctuations and crystallize a ranking in popularity that is misaligned with that based on quality, severely undermining the wisdom of crowd and producing giant distortions in the relative success of products.

The notion that popularity begets popularity is ubiquitous. It is generally known as Matthew effect, a term
introduced by R. K. Merton~\cite{merton_matthew_1968} to describe the disparity of recognition attributed to known
and relatively unknown scholars for producing work of comparable quality.
The issue of the possible distortions introduced by popularity has been studied even more directly by Salganick and collaborators. In their music lab experiment ~\cite{salganik_experimental_2006} they divided participants in non-communicating groups, and asked them to select songs from a menu common across groups.
When exposed to other members choices, groups produced very different popularity rankings due to the \lq\lq market" enhancing the idiosyncratic initial choices of few via the popularity driven dynamics. This supports the evidence that success is very hard to predict and engineer~\cite{watts_everything_2011}.

The problem becomes harsher when popularity bias compounds with the limited attention that users have to make quality discriminations. Evaluating products comes with a cognitive cost for the consumer that, in turn, influences
the amount of \emph{attention} he$/$she puts in this effort.
Previous research has shown
that the interplay between popularity bias that could be introduced by recommender systems
and consumer attention strongly impacts the quality of items sale ranking. Although traditional evaluation metrics of novel collaborative filtering recommender systems~\cite{10.1145/963770.963772} take into account such interplay, the problem is still considered hard to be solved.
Qiu \emph{et. al.}~\cite{qiu_limited_2017} studied a system of social recommendation where agents' choices depend both on items quality and
popularity among neighbors. They found \emph{i}) a very non-linear relationship between popularity and quality of items, \emph{ii})
a poor alignment between the quality and the popularity ranking, and \emph{iii})  an inverse relationship between said alignment and diversity
of the market.
Ciampaglia \emph{et al.}~\cite{ciampaglia_how_2018} studied a market where a combination of choice by quality and popularity is performed and showed
the existence  of a regime where  top-quality items are pushed at the top of the sale ranking.

The purpose of this paper is to better understand, from an analytical perspective, the interplay between popularity and quality as determinants of users choices in an idealized online cultural market, and the conditions that realize a desirable alignment between items ranking in popularity and quality.

For simplicity, we will not consider personalized recommendations, i.e.,
recommendation lists tailored to the specific user, or enhanced, say, by
collaborative filtering techniques. There are, indeed, scenarios in which
such personalization is not really needed, for example because the user
has already issued a query for a specific category of objects
matching her/his interests. For example, consider a user searching for books on the "Python programming language" on an e-commerce platform: the platform might have a few tens of books in its catalog about the Python language, and must decide a way to present them to the user, in a list more or less biased by popularity, knowing that the average user will inspect only
a subset of the items, in the case of a long list.
This scenario is different from the case of users
who just log in the platform without having specific ideas about what to buy/watch,  thus receiving personalized recommendations based on her/his navigation history, or history of similar users.

Moreover, for the sake of generality and of analytical tractability our approach assumes a generic and idealized online cultural market, and it is not meant to propose any specific
technique to be incorporated into a real recommendation system.
Nevertheless, we suggest potential applications of our findings. In particular, our results can help define the regimes in which a more or less aggressive dependence on popularity in the recommendation algorithm can have the undesired effect to promote items that are not those that qualify as top quality according to the average user
perception.\footnote{In Example \ref{example:application}, at the end of Section \ref{sec:popularity-pre-selection}, we provide a numerical example of application of our findings in a concrete scenario.}

Our contributions can be summarized as follows.
\begin{itemize}
\item We propose a simple model of an idealized online cultural market, in which $N$ items
have a hidden, intrinsic quality metric. Users pay attention to just a random subset of the items,
but are able to select the best-quality item within this subset. Items are recommended to users
by a ranking algorithm possibly biased by the current items' popularity.
\item We analyze the asymptotic system behavior as the number of user interactions grows large,
with the goal of understanding whether the ranking associated to quality eventually emerges,
aligning quality with popularity.
\item We discover that the system can have multiple equilibria depending on the ranking algorithm (parameterized
by a power-law exponent $\alpha \geq 0$) and the quality discrimination power of the users (parameterized
by the number $K$ of inspected items). Notably, different equilibria produce misalignments
which affect first top-quality items, before involving also lower-quality ones.
\item We characterize the minimum user discrimination power $K_{\min}$ that guarantees that
the desired alignment between quality and popularity is always achieved, in the long run.
\item We extend the model to multiple user classes, to account for the fact that different users can
perceive a different quality in the same item, and we conduct a preliminary investigation within this context.
\end{itemize}

Our main findings can be summarized as follows.
\begin{itemize}
\item In the special case in which items inspected by users are selected uniformly at random ($\alpha = 0$),
a minimum discriminating power $K_{\min} = 2$ allows alignment of quality and popularity,
at the expense of a low average quality of items selected by users.
\item In the case of ranking algorithms biased by popularity ($\alpha > 0$), there is
a harmful regime of mild popularity bias ($0 < \alpha \leq 1$) in which quality struggles to emerge, requiring
a disproportionately large $K$ to guarantee the desired alignment.
\item With stronger popularity bias ($\alpha > 1$), $K_{\min}$ is bounded and typically
small. Increasing $\alpha$, $K_{\min}$ approaches again the value of $2$, at the expense of increasing
unfairness among items (the top-quality item monopolizes the market).
\item An increasing fraction of na\"ive users, who deterministically select the most popular item,
makes things worse and worse, up to the point that alignment cannot be restored
by the other users, no matter their quality discrimination power.
\end{itemize}

The paper is organized as follows: Section \ref{sec:system} describes our model of online
cultural markets and our main assumptions. In Section \ref{sec:random-pre-selection}
we analyze the case where items inspected by users are selected uniformly at random, whereas Section \ref{sec:popularity-pre-selection}
deals with ranking algorithms biased by popularity. Section \ref{sec:fairness} deals with the
problem of parameter optimization to achieve a desired level of
fairness among the items. In Section  \ref{sec:multiclass}
we extend the model to the case of multiple user classes.
We discuss related work in Section \ref{sec:related}. At last in Section
\ref{sec:discussion} we discuss our contributions and findings,
point out the limitations of our approach, and
outline directions of future research.

\section{System model and main assumptions}
\label{sec:system}

We consider a fixed set $\mathcal N$ of $N = |{\mathcal N}|$ items that the recommender system/search-engine/e-commerce platform/social network
can offer to users issuing a certain query/interested in a given object category/exploring a given topic.
We assume that each item $i$, $i = 1,2,\ldots,N$, has an intrinsic quality $q_i \in \R$, which is unknown
to the system (and to the users). The notion of quality --- although the term is frequently (informally) used --- is subtle and hard to define.
For the purpose of our analysis, quality is simply an (unobserved) measure
that underlies a ranking. The ranking is the one upon which users would
converge upon they had the chance to chose items independently, i.e. in absence of social influence in general
and of algorithmic popularity bias in particular.
We assume that both the recommender system and the users are eventually interested in
promoting/discovering items having higher values of this hidden metric.
The absolute values of $q_i$ are not important, provided
that they produce the same ranking.
Hence, without loss of generality, we assume that items are indexed in increasing order of their quality, i.e., $q_1 < q_2 < \ldots < q_N$. Here item $N$
has the highest-quality and item $1$ the lowest.
The item ranking is assumed to be fixed over the time period
on which we study popularity dynamics.

Items also have an integer-valued \lq popularity weight' $w_i \in \N^+$, which represents, e.g., the number of times
they have been purchased/selected by the users, the number of views/likes/comments received, etc. For each item, this number increases over time,
when the item \lq wins' the competition with the other items.

Let $n$ be the number of competition rounds performed so far within the system
among the $N$ items, and $w_i[n]$ the weight accumulated by item $i$ after
the $n$-th competition.
For simplicity, we will assume that weight $w_i$ just counts
the number of competitions won by item $i$, i.e., it is increased by one each time
item $i$ wins. At time step $n$, a user receives a {\em randomized} list of recommended items drawn from the available set of $N$ objects.
The randomization introduced by the system is biased by the current popularity weights $w_i[n]$ of the objects (see next).
Due to limited time/attention, the user inspects only $K$ objects, and
selects the one with the best-quality among the $K$. By so doing, we account for the fact that users do not have enough time/skills to discover
the best-quality item among all alternatives, but they are at least able to do
so on a restricted set.
In the following we will refer to the subset of $K$ items
inspected by the user as {\em pre-selection}. If the user arriving at time $n$ selects objects $j$, popularity of $j$ increases by one\footnote{We assume for simplicity that competitions occur sequentially one after the other. In real systems many competitions can take place concurrently as several users interact at the same time with the online platform, but we neglect such micro-scale effects.}: $w_j[n] = w_j[n-1]+1$.

Popularity weights start from arbitrary initial values $w_i[0] > 0$.
For example, we can assume that at the beginning all items are equally popular with weight 1,
$w_i[0] = 1$, $i = 1 \ldots N$. Most of the results in this paper do not depend
on the set of initial weights, and we will point out explicitly when they do.

We denote by $b_i[n]$ the probability that item $i$ wins the $n$-th competition.
Such probability depends crucially both on the way in which the system presents available
items to the user (all of them or just a subset of them), for example through
a vertical scrollable list, and on the user behavior,
especially her/his patience to explore the alternatives and her/his ability to identify higher-quality
items, as expressed by discrimination parameter $K$.
Probabilities $b_i[n]$ allow us to define the concept of \emph{average quality index} of the online cultural market after $n$-th competition as $\overline{q}[n] = \sum_{i=1}^N i\,{b}_i[n]$.

Our main interest is to investigate what happens, in the long run (as $n \rightarrow \infty$), to popularity weights $w_i[n]$ and to average quality index $\overline{q}[n]$ as we vary the quality-discrimination parameter $K$. In particular, will asymptotic weights $w_i[n]$ be ordered according to the intrinsic quality of items? which is the minimum quality-discrimination power (i.e., the minimum $K$) requested to the users so that the best-quality item will eventually emerge? How does the  average quality index depend on the quality-discrimination power?

To answer these questions, we need to specify also the other fundamental ingredient of the model,
that is the law by which the restricted, random set of $K$ items is chosen from set $\mathcal N$,
which must reflect both the way items are internally promoted by the system and visually presented to the user,
and the additional randomness introduced by the user interaction with the online platform.

Similarly to previous work, we adopt the following popularity-biased {\em ranking model}. The ranking model was introduced in the context of an abstract model of network growth ~\cite{fortunato_2006} and used in a context similar to the present one in ~\cite{ciampaglia_how_2018}. It  aims at describing how items are selected by a ranked list, and it assumes that the entity performing the selection appreciates the differences in ranking, although not necessarily the quantity that underlies the ranking. Let $r_i[n] \in \{1 \ldots N\}$ be the
rank of item $i$ at the beginning of round $n$, in terms of popularity weight (i.e., $r_i[n]$ is the number of items - including $i$ -
in set $\mathcal N$ whose weight is at least as large as $w_i[n]$). Considering ranks, instead of absolute values $\{w_i[n]\}_i$, is a standard technique to avoid disproportionate bias towards items which have accumulated too much weight w.r.t. to others (for example because they have stayed in the system for much longer time). Then, we assume that $K$ items are successively drawn from set $\mathcal N$ (or the set of remaining items after previous extractions) with probability proportional to $r_i[n]^{-\alpha}$, where $\alpha \geq 0$ is an exponent reflecting the bias of the system/user towards popularity. The extreme values are $\alpha = 0$, corresponding to a system in which the $K$ items inspected by the user are chosen uniformly at random, irrespective of their popularity, and $\alpha \rightarrow \infty$, corresponding to a system in which
users deterministically focus their attention only on the current top-$K$ items in terms of popularity. The ranking model naturally assumes a monotonically decreasing probability to select an item as function of its popularity rank. While the power-law form of such
dependency is somewhat arbitrary, it is general enough to allow (via the parameter $\alpha$) to gauge how biased towards the most popular items is the preselection mechanism operated by the putative recommender system.

A simple, concrete example of mechanism described by the above model could be the following.
The system, in response to the user query, generates a random permutation of the $N$ items, according to the above power law
of the rank, and presents them to the user in a scrollable list; the user, because of
its limited budget of attention, explores only the first $K$ of the list, and after careful inspection
is able to select the one with the best quality.
Note that our model is not limited to this simple mechanism. More in general,
we can describe, through a single parameter $\alpha$, the generic bias towards popularity
of the restricted, random set of items explored by the user as a result of its interaction
with the platform (i.e., the items inspected by the user might not necessarily be the first $K$ of the list proposed by the system).

Since the systems we aim to describe are highly heterogeneous in terms of user behavior, we will consider several extensions to the base setting. In particular:
\begin{itemize}
\item we consider a case in which the parameter $K$ is not the same for all users to model user diversity
in terms of attention/discrimination power. This is achieved by introducing a probability distribution $\{p_k\}_k$ over the
quality-discrimination parameter $K$, which becomes an i.i.d. random variable across different users.
\item we also consider the case in which a fraction $f_m$ of users are too ingenuous/impatient and simply select the item which is currently the most popular.
$\mathcal N$ (some platforms indeed mark such item with a special flag, such as \lq\lq best seller").
We will refer to such users as na\"ive users. We finally consider the case in which users differ in term of preferences: quality is highly subjective and can be perceived differently by users.
We assign users to one
of a finite number $C$ of classes, according to fixed (generally heterogeneous) probabilities $\{f_c\}_{c=1}^{C}$.
Users belonging to the same class equally rank the quality of the $N$ items.
\end{itemize}

Table \ref{tab:notation} summarizes the paper notation.
\begin{center}
\begin{table}[htb]\centering
\boxed{
\begin{tabular}{l|l}
 symbol & description \\ \hline
  $N$ \rule{0pt}{3ex} & number of items \\
 $n$ & number of competition rounds \\
 $w_i[n]$ & popularity weight of item $i$ after round $n$ \\
 $\tilde{w}_i[n]$ & normalized popularity weight of item $i$ after round $n$ \\
 $r_i[n]$ & popularity rank of item $i$ at beginning of round $n$ \\
 $b_i[n]$ & winning probability of item $i$ at round $n$ \\
 $b_i = \lim_{n \rightarrow \infty} b_i[n]$ & asymptotic winning probability \\
 $\overline{q}[n]$ & average quality index at round $n$\\
 $\overline{q} = \lim_{n \rightarrow \infty} b_i[n]$ & asymptotic average quality index \\
 $\tilde{b}_i \sim (N-i+1)^{-\beta}$ & desired winning probability of item $i$ \\
 $\tilde{q} = \sum_{i=1}^N i\,\tilde{b}_i$ \rule{0pt}{3ex} & desired average quality index \\
 $\alpha$ & pre-selection power-law exponent \\
 $\beta$ & fairness power-law exponent \\
 $K$ & quality discrimination power of users \\
 $p_k$ & distribution of $K$ (if used) \\
 $f_m$ & fraction of na\"ive users \\
 $C$ & number of user classes \\
 $f_c$ & probability the a user belongs to class $c$
\end{tabular}
}
\caption{Notation}\label{tab:notation}
\end{table}
\end{center}

\subsection{Toy example}\label{subsec:toyexample}
\begin{figure}[htb]
\centering
\includegraphics[width=8cm]{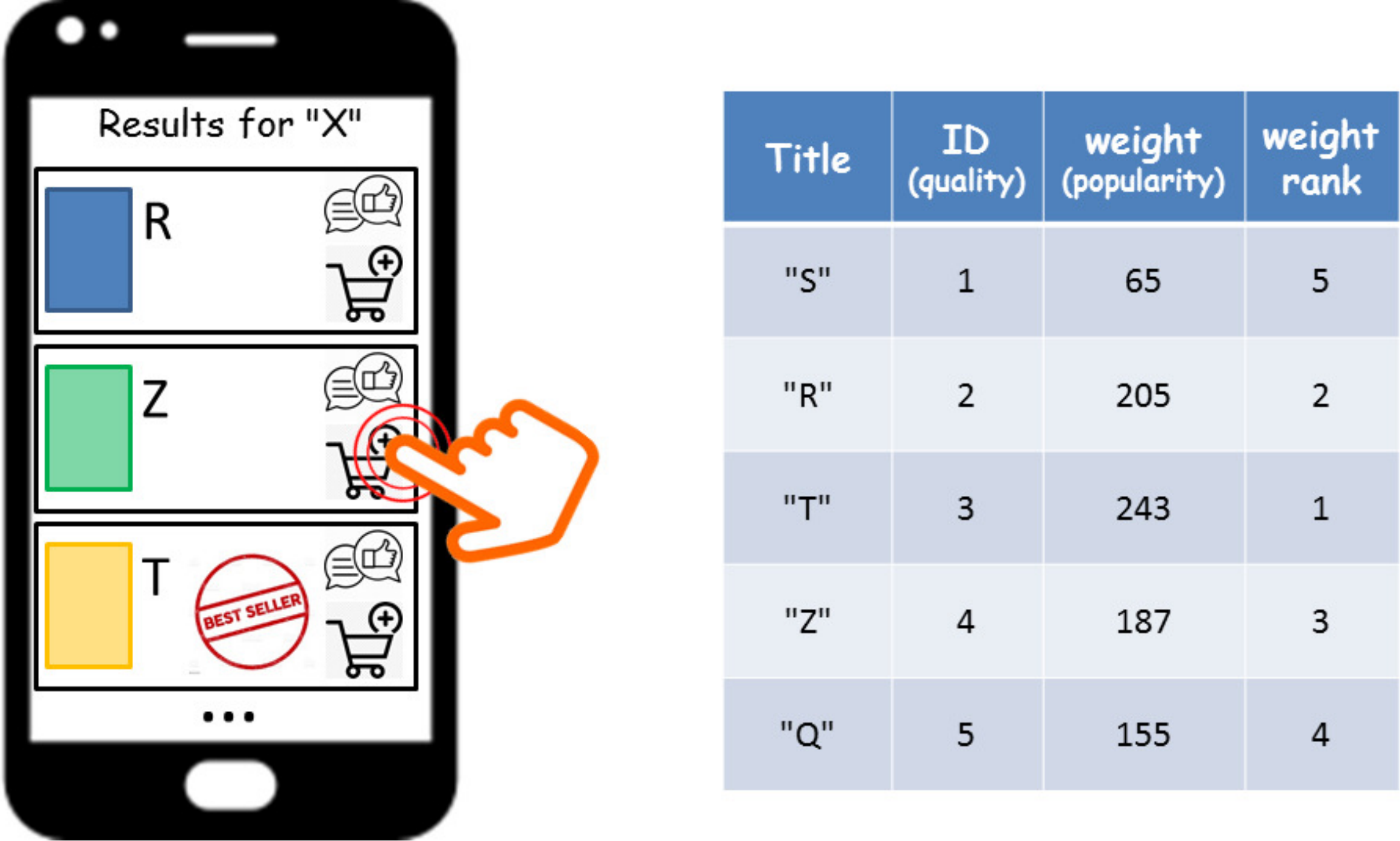}
\caption{Illustration of competition among $N=5$ items, with $K = 3$ items viewed (and inspected) by a certain user:
instance of user view (left) and item features (right table). The stylized hand indicates the click over item $Z$
which will be selected by a non-na\"ive user, being the highest-quality item among the inspected 3 items.\label{fig:esempioK}}
\end{figure}

For clarity, we present a numerical example of a competition round
 in a system operating under the base setting described above, see Figure \ref{fig:esempioK}.
Suppose that a user searches for books about some topic \lq\lq X" on an e-commerce site,
and that 5 books match the query. Books 1,2,3,4,5, ordered according to increasing
intrinsic quality, have title \lq\lq S",  \lq\lq R",  \lq\lq T",  \lq\lq Z",  \lq\lq Q",
respectively, and their current popularity rank is 5,2,1,3,4, respectively, computed for example
on the current number of purchased copies, which provides weights $w_i[n]$ at the beginning
of the $n$-th competition (see right table on  Figure \ref{fig:esempioK}).
Suppose the user has time to inspect the reviews of the first 3 books of the
list shown to her/him (left view in Figure \ref{fig:esempioK}),
discovering the one with the best quality among titles \lq\lq R",  \lq\lq Z",  \lq\lq T",
which happens to be \lq\lq Z". This particular (sub)list, in the case
of a rank-based power-law pre-selection with exponent $\alpha = 1$, is
generated by the recommender system with probability:
$$ \frac{\frac{1}{2}}{1+\frac{1}{2}+\frac{1}{3}+\frac{1}{4}+\frac{1}{5}} \cdot
 \frac{\frac{1}{3}}{1+\frac{1}{3}+\frac{1}{4}+\frac{1}{5}} \cdot
 \frac{1}{1+\frac{1}{4}+\frac{1}{5}} \approx 0.028.$$

The formula above is the product of the three terms that correspond
to the probabilities to select R first, Z second and T third. Each of this probabilities is proportional to the
inverse of the current rank of the item ($\alpha =1$ in the example). This accounts for the numerators of the three terms.
The denominators are the needed normalization to reflect the fact that once an item has been selected, it cannot
be selected again. So, for example, once R (rank=2) has been selected, only 4 items are still available, those
of rank 1,3,4,5, which explains the denominator of the second term.

Note that, in the case $f_m = 0$, the user will end up buying book \lq\lq Z", which has the highest quality among the 3 inspected books.
Otherwise, with probability $f_m$ the user is a na\"ive one, who would instead decide to buy
the best-seller \lq\lq T".

\section{Uniform pre-selection}
\label{sec:random-pre-selection}
As a warm-up, we start considering the simple case $\alpha = 0$, where any combination of $K > 1$ (distinct) objects
has the same probability to be pre-selected.
In this case, the winning probability $b_i$ of item $i$ does not depend on $n$ and reads:
\begin{equation}\label{eq:birandom}
b_i = \begin{cases}
0 \qquad & i < K\\
\frac{\binom{i-1}{K-1}}{\binom{N}{K}} \quad & i \geq K
\end{cases}
\end{equation}
since it is equal to the fraction of combinations in which
item $i$ is present, and the other $K-1$ elements of the pre-selection
are chosen from the set of $i-1$ lower-quality items.

The above winning probabilities, for $i \geq K$, are positive and strictly increasing with index $i$.
Moreover, they do not depend on current weights $w_i[n]$. Therefore,
${\bm w}[n] = (w_1[n], w_2[n], \ldots , w_N[n]) \sim {\bm w}[0] + \textrm{Multi}(b_1, b_2, \ldots , b_N; n)$, i.e. ${\bm w}[n]$
follows a multinomial probability distribution with parameters $(b_1, b_2, \ldots , b_N; n)$,
shifted by initial weights ${\bm w}[0]$.
Let $E_n$ be the event that, after $n$ competitions, weights associated to items with positive winning probability
are correctly ordered according to the
quality of items, i.e., $$E_n = \{w_K[n] < w_{K+1}[n] < \ldots < w_N[n]\}.$$

Exploiting classic concentration results we can prove the following:
\begin{proposition}[Asymptotic weights with increasing winning probabilities]\label{prop:increasing}
When non-null winning probabilities $\{b_i\}_i$ are strictly increasing with index $i$,
as the number of competitions $n$ grows to infinity, the associated weights are almost surely
correctly ordered, i.e.,
$$ \PP[\lim_{n \rightarrow \infty} E_n] = 1.$$
\end{proposition}
\begin{proof}
See Appendix \ref{app:increasing}.
\end{proof}

The above result suggests that a minimum discrimination effort on behalf of the users (i.e., $K=2$)
is sufficient, in the case $\alpha = 0$, to eliminate the effects of popularity bias
and guarantee, in the long run, that high-quality items will eventually emerge as popular, with item popularity
perfectly aligned with item quality. Note that in the case of $K=1$, instead, quality never comes into play
in determining selections, hence any permutation of the items is equally likely
to provide the final ranking of the items in terms of popularity.

The above results can be easily extended to the case of heterogeneous users
characterized by a distribution $\{p_k\}_k$ of the discrimination
parameter $K$. Indeed, denoting by $b_{i}(k)$ the dependency of \equaref{birandom} on $K$, we simply have
the weighted average:
$$ b_i = \sum_{k \geq 1} p_k b_{i}(k).$$
Note that sequence $\{b_i\}_i$ is still strictly increasing for all distributions
$\{p_k\}_k$ except the one in which $p_1 = 1$ (i.e., fixed $K=1$), hence
Proposition \ref{prop:increasing} still applies except for this degenerate case.

At last, considering for simplicity the case of constant $K$,  we examine the impact of $f_m$,
the fraction of na\"ive users who always select the most
popular item. Denoting by $\hat{b}_i$ the winning probability of item $i$ with $f_m > 0$,
we have $$\hat{b}_i = f_m \cdot I_{i = \argmax_j w_j} + (1-f_m) b_i$$
where $I_{i = \argmax_j w_j}$ is the indicator function of the event that item $i$ is currently
the most popular, while $b_i$ is the winning probability for non-na\"ive users.
In this case the most popular item is not guaranteed to be item $N$.
Indeed an item $i^* < N$ can become the most popular, provided
that its winning probability is higher than that of all other items, and most crucially
of that of item $N$, i.e., provided that:
$$ f_m + (1-f_m) b_{i^*} > (1-f_m) b_N \Rightarrow f_m > \frac{b_N - b_{i^*}}{1 + b_N - b_{i^*}}$$
which means that, if $f_m$ is large enough, item $i^*$ can stably occupy the most popular position.
As we increase $f_m$, starting from 0, we see that $N-1$ is the first item that can replace $N$
on top of the list. Since configurations in which item $N$ is not the most popular are undesirable,
we can compute the maximum value of $f_m$ such that
only item $N$ can be the most popular. Indeed, since \mbox{$b_{N} - b_{N-1} = \frac{K(K-1)}{N(N-1)}$},
we obtain the condition:
$$ f_m < \frac{K(K-1)}{N(N-1)+K(K-1)}.$$
We observe that, when $K \ll N$, a small fraction of na\"ive users
is enough to (potentially) disrupt the optimal configuration in which
item $N$ is the most popular.

Having analyzed the simple case of uniform pre-selection, in next section we move on to
investigate what happens in the more challenging case of popularity bias
in the pre-selection of items, i.e., when current popularity weights $w_i[n]$
are used by the system to build the list of items recommended to the user.

\section{Popularity-based pre-selection}
\label{sec:popularity-pre-selection}
When $\alpha > 0$, more popular items are more likely to be proposed to the user,
and if they win the competition (which depends on their intrinsic quality) they become even more popular.
This endows the system with a self-reinforcing property reminiscent of the rich-get-richer phenomenon.
What happens in the long run? Will item popularity reflect the intrinsic item quality, or can the process
lead to (undesirable) configurations in which the most popular item is not the best?

Our system can be modeled as a special case of P\'{o}lya urn \cite{polyabook} with $N$ colors ,
where $w_i[n]$ is the number of balls of color $i$ after the $n$-th round.
At competition round $n$, $K$ balls are selected from the urn, and a new ball
of the winning color is added to the urn.
In contrast to classic P\'{o}lya urn models, the analysis here is complicated by the fact
that the selection of $K$ balls is a complex function of the current rank of the $N$ colors
in terms of their ball count.
Indeed, recall that our system, mapped onto a P\'{o}lya urn, works as follows: we start with an empty set $\mathcal S$ of
pre-selected balls; at each of $K$ iterations, a ball of color $i$ is chosen with probability
proportional to $r_i[n]^{-\alpha}$, where $r_i[n]$ is the rank of color $i$ at the beginning of the round,
and added to set $\mathcal S$. If $j$ is the highest-quality color in set $\mathcal S$, a ball
of color $j$ is added to the urn at the end of the competition: \mbox{$w_j[n] = w_j[n-1] + 1$},
whereas \mbox{$w_i[n] = w_i[n-1]$} for all $i \neq j$.

Moreover, we may or may not allow the repetition of colors (items) in the set of $K$ balls.
This distinction leads to two different pre-selection schemes, that we will call
{\em with-item-repetition} and {\em without-item-repetition}.
As the name suggests, {\em with-item-repetition} means that the $K$ balls are independently selected with a
probability that depends only on the ranking of the colors at the beginning of the round.
Specifically, each of the $K$ selected balls belongs to color $i$ with probability
\begin{equation}\label{eq:piwith}
p_i[n] = \frac{r_i[n]^{-\alpha}}{\sum_j r_j[n]^{-\alpha}}.
\end{equation}
Note that, by so doing, we could obtain more than one ball of a given color in our pre-selection.

In {\em without-item-repetition} pre-selection, we enforce that $K$ balls of {\em distinct}
colors are selected by re-normalizing the probabilities to choose a ball of a given color
among the remaining colors. Let $p_{i,k}[n]$ be the probability to select the $k$-th
($k = 1,\ldots,K$) ball among those of color $i$, and ${\mathcal C}_k$ the set of colors
extracted in the first $k$ choices, starting with ${\mathcal C}_0 = \emptyset$.
We have:
\begin{equation}\label{eq:piwithout}
p_{i,k}[n] = \begin{cases}
0 \qquad & i \in {\mathcal C}_{k-1} \\
\frac{r_i[n]^{-\alpha}}{\sum_{j: j \notin {\mathcal C}_{k-1}} r_j[n]^{-\alpha}} \qquad & i \notin {\mathcal C}_{k-1}.
\end{cases}
\end{equation}

The {\em without-item-repetition} pre-selection scheme more realistically describes a system that
offers $K$ distinct alternatives to the user (see example in Section \ref{subsec:toyexample}).
Unfortunately, this scheme is more difficult to analyze than {\em with-item-repetition}, as it requires
to consider all dispositions of $N$ elements over $K$ positions.
Conversely, {\em with-item-repetition} is easy to analyze, but
it's less realistic, since items can appear multiple times in the pre-selection.
After extensive numerical experiments, we discovered that almost all properties
of the system operating under {\em without-item-repetition} are the same as
those of the system operating under {\em with-item-repetition}. Moreover, as we will see,
analytical results for the {\em with-item-repetition} scheme are very close to those numerically obtained
for {\em without-item-repetition}, especially when $K \ll N$.
For these reasons, in the following we will mainly focus on the {\em with-item-repetition} scheme.

We start with a general result valid for a generic pre-selection scheme
based on the popularity rank of the items.
Let $\tilde{w}_i$ be the asymptotic fraction of balls of color $i$ in the urn as the number of
competitions tends to infinite:
$$ \tilde{w}_i = \lim_{n \rightarrow \infty} \frac{w_i[n]}{\sum_j w_j[n]} =
\lim_{n \rightarrow \infty} \frac{w_i[n]}{n + n_0} $$
where $n_0$ is the initial number of balls in the system.

Let $B: [0,1]^N \rightarrow [0,1]^N$ be the function that, given
a vector ${\tilde{\bm w}}$ of normalized weights, provides the
corresponding vector ${\bm b}$ of winning probabilities.
We will restrict ourselves to functions $B(.)$ that depends
on ${\tilde{\bm w}}$ only through the rank of the items.

\begin{assumption}[rank-based winning probabilities]\label{ass:B}
Function $B: [0,1]^N \rightarrow [0,1]^N$ is an injective function
of the rank of weights  ${\tilde{\bm w}}$.
\end{assumption}

\begin{definition}[system stable point]\label{def:stablepoint}
A system stable point is a stochastic vector ${\bm e} \in [0,1]^N$ (a vector with non-negative entries that add up to one),
which is a fixed point of function $B(.)$, i.e., $B({\bm e}) = {\bm e}$, and, in addition,
$e_i \neq e_j$ for $i \neq j$.
\end{definition}

\begin{proposition}[asymptotic system behavior] \label{prop:fixed-point}
Under Assumption \ref{ass:B}, as \mbox{$n \rightarrow \infty$},
normalized weights $\tilde{\bm w}[n]$ converge to a system stable point.
\end{proposition}
\begin{proof}
See Appendix \ref{app:fixed-point}.
\end{proof}

Note that the system can have multiple stable points. Each stable point is associated to a
particular permutation of the $N$ items. For this reason, we will also say that a permutation
is stable if it is associated to a system stable point.

We will call desirable equilibrium the stable point associated to the natural
permutation $1,2,\ldots,N$ in which item popularity is perfectly aligned with item quality.
We will call instead spurious equilibrium a stable point associated to
a permutation different from the natural one.

The number and identity of stable points clearly depends on function $B(.)$.
In the {\em with-item-repetition} scheme, we have a simple expression
for winning probabilities $\{b_i\}_i$.
Indeed, denoting by $s_i = \sum_{j \leq i} p_i$ the cumulants of item selection probabilities \equaref{piwith}
(we have dropped the dependency on $n$ for simplicity), we can write, for $i > 1$:
\begin{equation}\label{eq:birep}
b_i = s_i^K - s_{i-1}^K
\end{equation}
(for $i=1$, we have the special case $b_1 = p_1^K$).

Equation \equaref{birep} can be understood as follows: for item $i$ to be the highest-quality
item selected among $K$ independent choices with repetition we need that no higher-quality
object has been selected for $K$ times (term $s_i^K$); this event includes however also the event
that color $i$ has never been selected among the $K$ choices (equivalently, objects
of index up to $i-1$ have been chosen $K$ times), hence the probability $s_{i-1}^K$ of this event
has to be subtracted from previous event.

In the {\em without-item-repetition} scheme, instead, the evaluation of $\{b_i\}_i$
requires, unfortunately, to enumerate all possible ordered sequences $\sigma_i$ of $K$ objects,
containing item $i$ and other $K-1$ lower-index items (without repetition):
\begin{equation}\label{eq:bidis}
b_i = \sum_{\sigma_i} \prod_{\substack{k=1,\ldots,K \\ j = (\sigma_i)_k}} p_{j,k}
\end{equation}
where $p_{j,k}$ are given by \equaref{piwithout}.
In this case the computation of $\{b_i\}_i$ is feasible only for small values of $N$, $K$.

\begin{figure}[htb]
\centering
\includegraphics[width=8cm]{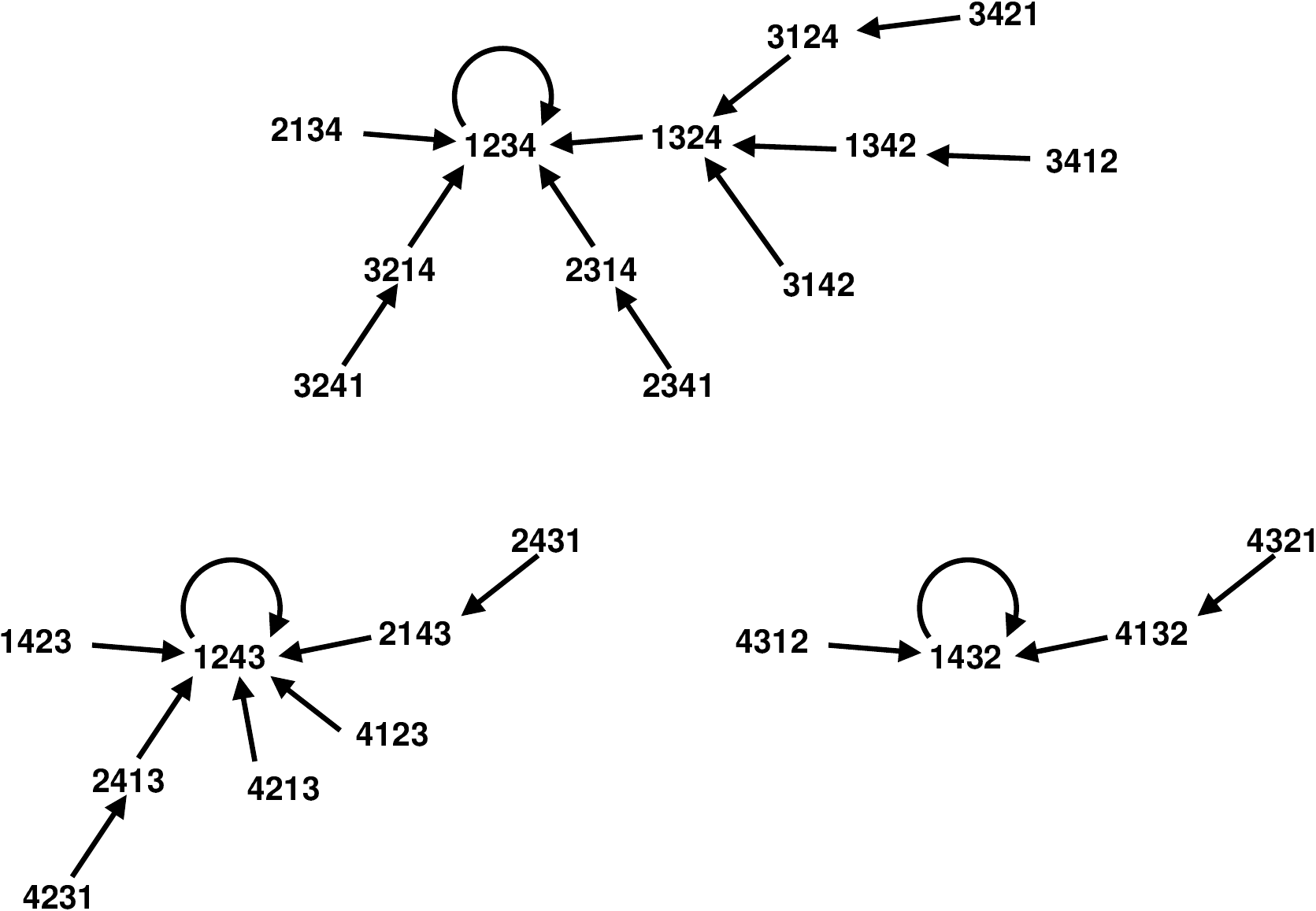}
\caption{Permutation graph for the system $N = 4$, $K=2$, $\alpha = 1$, {\em with-item-repetition}.\label{fig:graph42}}
\end{figure}

To find out all of the stable points of a system, we have followed different approaches.
For small systems (say $N < 20$) one can simply test all $N!$ permutations,
and check whether each of them is stable (according to Definition \ref{def:stablepoint}).
To better illustrate the system dynamics, one can build, for a given choice of $N,K,\alpha$
the {\em permutation graph}, containing $N!$ nodes, one for each permutation, and add a
direct edge from node $i$ to node $j$ if function $B(.)$ maps permutation $i$ into permutation $j$.
Figure \ref{fig:graph42} shows the permutation graph for the system $N=4$, $K=2$, $\alpha = 1$, in the case
{\em with-item-repetition}. In this case we have $4! = 24$ nodes, and 3 fixed points of function $B(.)$,
associated to permutations $\{1, 2, 3, 4\}$, $\{1, 2, 4, 3\}$ and $\{1, 4, 3, 2\}$.
However, only $\{1, 2, 3, 4\}$ and $\{1, 2, 4, 3\}$ are stable points, because
with $\{1, 4, 3, 2\}$ we have $b_2 = b_3$. Detailed simulations of the corresponding P\'{o}lya urn
confirm that the actual system can only converge to the above two stable points.

The permutation graph can also be used to define the \emph{attractiveness} of
a given fixed point of function $B(.)$, i.e., of a permutation $f$ such that $B(f) = f$.
The attractiveness $a(f)$ of fixed point $f$ is defined as the size of the (weakly) connected
component containing $f$, divided by the number $N!$ of all permutations. Note that $\sum_{f: B(f) = f} a(f) = 1$.
For example, in the scenario of Figure \ref{fig:graph42} we have
$a(\{1, 2, 3, 4\})=\frac{1}{2}$, $a(\{1, 2, 4, 3\})=\frac{1}{3}$,
and $a(\{1, 4, 3, 2\})=\frac{1}{6}$.

For large systems, we cannot examine all possible permutations. To overcome this problem, one
approach is to perform a randomized search of stable points by starting from an arbitrary
stochastic vector $\tilde{\bm w}$, i.e., some random positive values $\{\tilde{w}_i\}_i$ summing up to 1, and iteratively apply
function $B(.)$ until we hit a fixed point. This procedure is described in details in
Algorithm \ref{alg:search}. In the permutation graph, this means to start from
a random node, and follow the direct path up to the node with the self-loop.
One problem of this approach is that we need to perform a number of searches large enough
to fall at least once in each connected component of the permutation graph, and therefore we are not
guaranteed to discover all of the stable points.

\begin{algorithm}[thb]
\begin{algorithmic}[1]
\Require $N \in \N^+$, $K \in \{1,\ldots,N\}$, $\alpha \in \R^+$
\State Choose random positive values \mbox{$\{\tilde{w}_i\}_{i=1}^{N}: \sum_{i=1}^{N} \tilde{w}_i = 1$} \Comment{Initialization of $w_i$'s}
\State $fixedpoint \Leftarrow \FALSE$
\While{($fixedpoint = \FALSE$)}
\State $r_i \Leftarrow$ index of $i$ in sorted $\tilde{\bm w}$ , $r_i \in \{0,\ldots,N-1\}$
\State $p_i \Leftarrow (N-r_i)^{-\alpha}, \forall i$ \Comment popularity of $i$
\State normalize $p_i$'s such that $\sum_{i=1}^{N} p_i = 1$
\State compute $b_i = \PP[\text{object} \,\,i\,\, \text{wins}], \forall i$  \Comment using either \equaref{birep} or \equaref{bidis}
\If{$\bf b = \tilde{\bm w}$}
\State	$fixedpoint \Leftarrow \TRUE$
\Else
\State $\tilde{\bm w}\Leftarrow \bm b$
\EndIf
\EndWhile
\If{$b_i \neq b_j, \forall i \neq j$}
\State $\bm b$ is a stable point
\EndIf
\end{algorithmic}
\caption{Randomized computation of equilibria}
\label{alg:search}
\end{algorithm}

As we will see, in most cases the only stable permutations are those that, starting from the natural
permutation, change the order of the top $M$ items, with $M \ll N$. Therefore, in practice
one can find all stable points by sequentially exploring all permutations generated in lexicographic order,
and stopping the search when no more stable points are found comprising swaps of the top $M+1$ items. The random search performed by Algorithm 1 is then to be considered a last resort, when
exact approaches become unfeasible ($N$ or $M$ too large). We emphasize that the above computational issue
pertains only the analytical prediction of system stable points, while the actual algorithm implemented by the recommendation system,
which produces just randomized lists biased by popularity (see toy example in Section \ref{subsec:toyexample}),
can be applied to arbitrarily large $N$ without
scalability problems, but being oblivious of where it is going to converge.

\begin{figure}[htb]
\centering
\includegraphics[width=9cm]{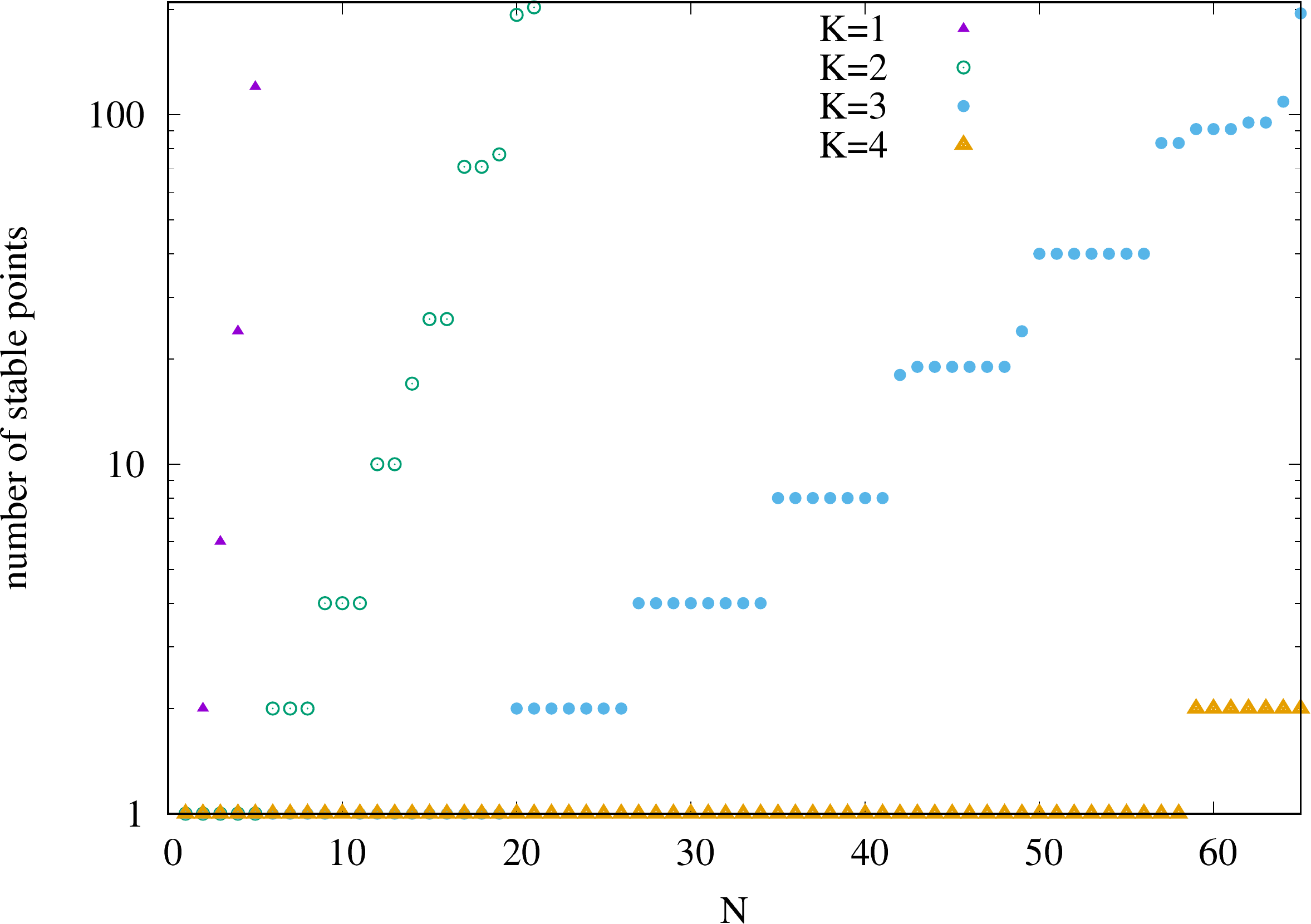}
\caption{Number of stable points for {\em without-item-repetition}, $\alpha = 1$.\label{fig:stableinv1}}
\end{figure}

As another interesting example, consider the {\em without-item-repetition} scheme with $\alpha = 1$.
Figure \ref{fig:stableinv1} shows, on a logarithmic vertical scale, the number of
stable points as function of $N$, for $K = 1,2,3,4$.
Note that for $K = 1$ all $N!$ permutations are feasible.
As we increase $K$, the number of stable points is brought down drastically,
though it still increases exponentially for large $N$.

For example, in the case of $N = 10$ and $K = 2$, there are 4 stable points, corresponding
to permutations: $$\{1, 2, 3, 4, 5, 6, 7, 8, 9, 10\}, \{1, 2, 3, 4, 5, 6, 7, 8, 10, 9\},
\{1, 2, 3, 4, 5, 6, 7, 9, 8, 10\}, \{1, 2, 3, 4, 5, 6, 7, 10, 9, 8\}.$$
Notice that, in addition to the (desired) natural permutation,
other three spurious permutations can emerge in which some of the highest-quality
items are permuted (the top two or top three items).

Interestingly, for given $N$, if we take $K$ large enough we obtain a single
stable point. We will see that, when the stable point is unique, it corresponds to the
natural permutation, i.e., the desired equilibrium. For example, $K=3$ is enough to
obtain as stable point only the natural permutation for all values of $N$ up to 19.
As another case, $K=4$ is required to achieve only the desired equilibrium
with $N=50$ items.
In the next subsection we will precisely address the problem of
determining when a given system $(N,K,\alpha)$ can only achieve the desired equilibrium,
while no spurious permutations can emerge from the competition among items.

\subsection{Computation of $K_{\min}$}\label{subsec:kmin}
We define $K_{\min}$ as the minimum value of $K$
that admits the natural permutation $1,\ldots,N-1,N$ as the only stable permutation.
By virtue of the
following lemma, to compute $K_{\min}$ it is sufficient, in most cases, to determine whether the alternate permutation
$1,\ldots,N,N-1$, in which the two highest-quality objects are permuted, is stable, .

\begin{lemma}[critical permutation]\label{lemma:critical}
Under popularity-biased pre-selection, the minimum value $K_{\min}$ is determined by the stability
of the {\em critical} permutation $1,\ldots,N,N-1$.
\end{lemma}
\begin{proof}
See Appendix \ref{app:critical}.
\end{proof}

We first consider the case {\em with-item-repetition}, for which
$K_{\min}$ can be characterized exactly.
Suppose that the current permutation, induced by increasing weights, is the critical permutation $1,\ldots,N,N-1$.
Then we have $p_N = 2^{-\alpha}/G$, $p_{N-1} = 1/G$, where $G = \sum_{i=1}^{N} i^{-\alpha}$
is the normalization constant of the power-law popularity bias.
We also introduce the cumulants $s_N = 1$, $s_{N-1} = 1-p_N$, $s_{N-2} = 1 - p_N - p_{N-1}$,
with which we can write the winning probabilities of the top two objects as:
\begin{eqnarray*}
b_N &=& s_N^K - s_{N-1}^K \\
b_{N-1} &=& s_{N-1}^K - s_{N-2}^K.
\end{eqnarray*}
The critical permutation is not stable when \mbox{$b_{N-1} < b_N$}, i.e., when
\begin{equation}\label{eq:conduniq}
2 s_{N-1}^K < s_N^K + s_{N-2}^K.
\end{equation}
Substituting the cumulants, the above inequality becomes:
\begin{equation}\label{eq:leftright}
2 (1-G^{-1} 2^{-\alpha})^K \leq 1 + (1-G^{-1}-G^{-1}2^{-\alpha})^K
\end{equation}
which provides, as function of $G$ and $\alpha$, the condition that $K$ must satisfy
so that the critical permutation is not stable.
Therefore, by Lemma \ref{lemma:critical} we have:
\begin{equation}
K_{\min} \triangleq \min_K \left\{ K: 2 (1-G^{-1} 2^{-\alpha})^K \leq 1 + (1-G^{-1}-G^{-1}2^{-\alpha})^K \right\}. \label{eq:kminrep}
\end{equation}

The following proposition provides a summary of interesting properties
of $K_{\min}$ as function of parameters $N$, $\alpha$.
In particular, since the catalog size can in same cases
be very large, it is interesting to see what happens when $N \rightarrow \infty$,
though the item set is to be considered finite in any feasible scenario.

\begin{proposition}[Properties of $K_{\min}$]\label{prop:kmin}
For any given $N \geq 1$, $\alpha \geq 0$, there exists an integer value \mbox{$K_{\min} \geq 1$} such that
for any \mbox{$K \geq K_{\min}$} the only stable permutation is the natural one.
For fixed $\alpha$, $K_{\min}$ is a non-decreasing function of $N$.
As a special case, $K_{\min} = 2$ for $\alpha = 0$. When \mbox{$0 < \alpha \leq 1$},
$K_{\min} \rightarrow \infty$ as $N \rightarrow \infty$.
For fixed $\alpha > 1$, \mbox{$K_{\min} \rightarrow K_{\min}^\infty$} as $N$ increases,
where $K_{\min}^\infty$ is a constant which depends on $\alpha$.
\end{proposition}
\begin{proof}
See Appendix \ref{app:kmin}.
\end{proof}

In the following, we will see that properties of $K_{\min}$ listed in Proposition
\ref{prop:kmin} hold also under {\em without-item-repetition}.
The following result instead is specific to the case {\em with-item-repetition}:
\begin{corollary}\label{coro:bigalfarep}
For given $N>1$, in the case {\em with-item-repetition},
$K_{\min} \rightarrow \infty$ as $\alpha \rightarrow \infty$.
\end{corollary}

\begin{proof}
We can simply upper bound the right-hand side of \equaref{leftright}, for any $\alpha$ and $K$, by $2-G^{-1}$, obtaining
the weaker inequality:
$$ 2\left(1-\frac{1}{G 2^\alpha}\right)^K \leq 1 + \left(1-\frac{1}{G}-\frac{1}{G 2^{\alpha}} \right)^K \leq 2 - \frac{1}{G} $$
which is satisfied by
\begin{equation}\label{eq:newcoro}
K \geq \frac{\log(1-(2G)^{-1})}{\log(1-(G 2^{\alpha})^{-1})}
\end{equation}
The assert follows by noticing that the right-hand side of \equaref{newcoro} tends to infinity
as $\alpha \rightarrow \infty$.
\end{proof}

\begin{figure}[htb]
\centering
\includegraphics[width=9cm]{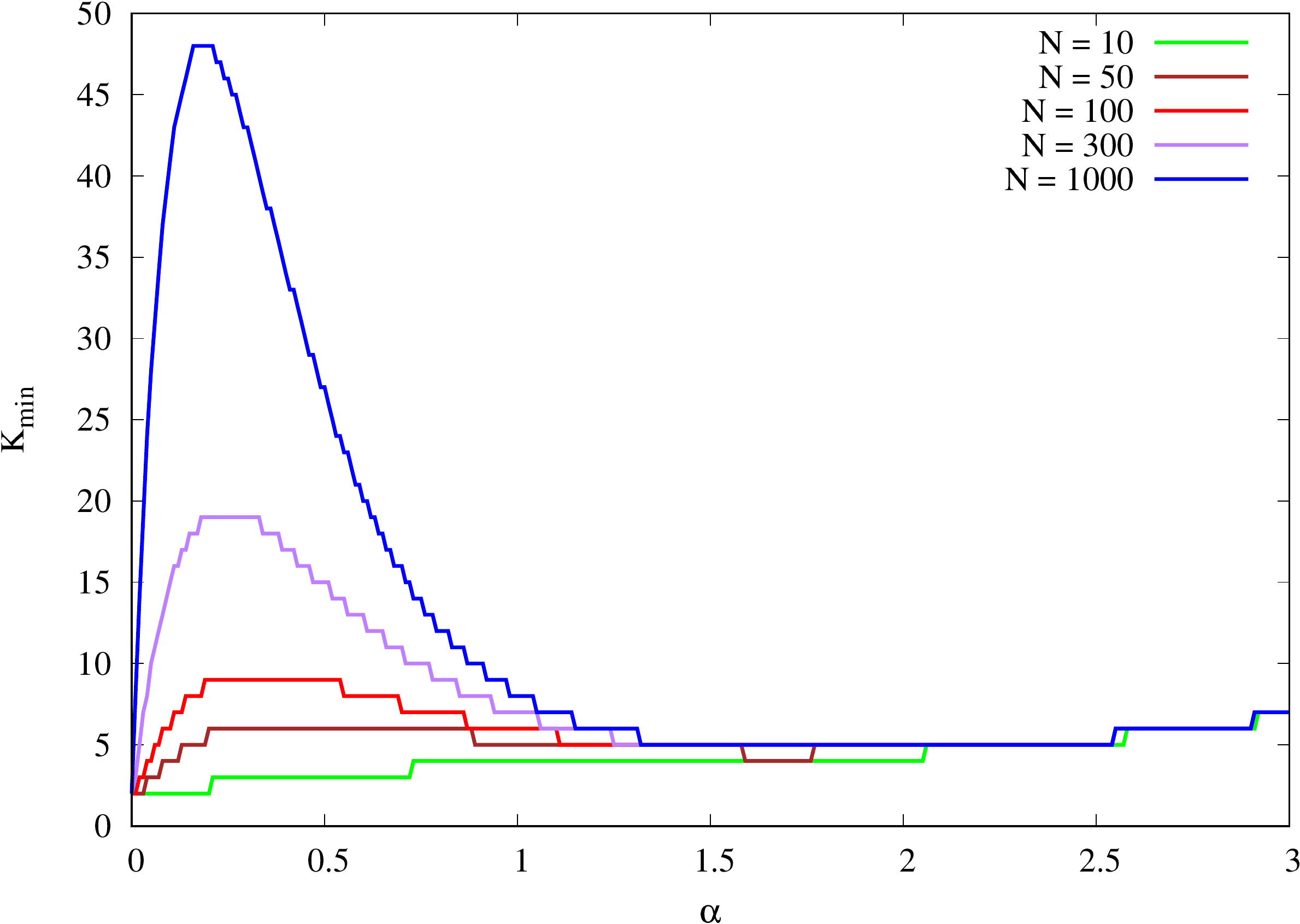}
\caption{Minimum value of $K$ such that the natural permutation is the only stable point,
for different values of $N$, {\em with-item-repetition}.\label{fig:minkalfa}}
\end{figure}

Figure \ref{fig:minkalfa} shows $K_{\min}$ as function of $\alpha$, for different
values of $N$, obtained by solving numerically \equaref{kminrep}.
Results in Figure \ref{fig:minkalfa} confirm the properties in Proposition
\ref{prop:kmin}. In particular, $K_{\min}$
becomes unbounded, as $N$ increases, for $0 < \alpha \leq 1$.
As predicted by Corollary \ref{coro:bigalfarep}, for large $\alpha$ the value
$K_{\min}$ (slowly) increases with $\alpha$.

\begin{remark}
When $K < K_{\min}$, the system is not guaranteed to produce, in the long run, the desired
alignment between quality and popularity, and can end up operating under
one of many suboptimal configurations containing severe misalignments in the top of the list.
The frequencies with which we can observe such undesirable operating points depend
crucially on the distribution ${\bm w}[0]$ of initial weights.
Extensive numerical simulations reveal that undesirable configurations are not at all \lq rare events',
and show up with frequency comparable to that of the natural configuration, even when
we start with all weights initially equal to 1. If we start with random initial weights, e.g., an i.i.d weight
for each item, we increase the likelihood to \lq crystallize' initial suboptimal configurations, and
correspondingly reduce the probability to achieve the desired alignment.
\end{remark}

\begin{remark}
Proposition \ref{prop:kmin} provides some fundamental insights into the behavior of popularity-biased
cultural markets. Specifically, it suggests there exists a harmful regime
of mild popularity bias ($0 < \alpha \leq 1$) in which desired alignment of popularity and quality
requires large quality discrimination power on behalf of users. Since
in this regime $K_{\min}$ diverges as $N$ increases, systems with large catalog size
are essentially unpredictable in this case, and can possibly produce severe misalignments among high-quality items.
Surprisingly, the special case $\alpha = 0$ (uniform pre-selection) behaves as a singularity,
requiring the minimum discrimination power $K_{\min} = 2$ for any $N$.
\end{remark}

\subsubsection{Extensions to heterogeneous discrimination power.\\}
Previous analysis of $K_{\min}$ can be extended to account for the presence of heterogeneous users with different
behavior. One easy extension is to consider a fraction $f_m$ of users deterministically choosing
the current most popular item, while the remaining fraction $1-f_m$ of users employ quality discrimination power $K$.
In this case, under the crucial permutation we have:
\begin{eqnarray*}
b_N &=& (1-f_m) (s_N^K - s_{N-1}^K) \\
b_{N-1} &=& f_m + (1-f_m) (s_{N-1}^K - s_{N-2}^K)
\end{eqnarray*}
and condition \mbox{$b_{N-1} < b_N$} becomes:
$$ f_m + (1-f_m)\left[2 \left(1-\frac{1}{G 2^{\alpha}}\right)^K-1 - \left(1-\frac{1}{G} - \frac{1}{G 2^{\alpha}}\right)^K\right] < 0.$$
We observe that the left-hand side is positive for $K=1$, and becomes negative for sufficiently large $K$ provided
that $f_m < 1/2$ (the term in square brackets tends to -1 as $K \rightarrow \infty$).
Therefore, \mbox{$K_{\min} \rightarrow \infty$} as \mbox{$f_m \rightarrow 1/2$}, and no $K_{\min}$ exists for $f_m \geq 1/2$.
With fixed $f_m < 1/2$, all properties stated in Proposition \ref{prop:kmin} are still valid, as one can verify along the same lines
of the proof reported in Appendix \ref{app:kmin}.
Moreover, $K_{\min}$ is an increasing function of $f_m$, which can be proven by applying the
implicit function theorem, and noticing that \mbox{$\frac{\partial F}{\partial f_m} = 1$} at the points in which
$F(x,f_m) = 0$ (see Appendix \ref{app:kmin}).

As another extension we can consider the case of users with i.i.d. values of $K$,
distributed according to the discrete law $\{p_k\}_k$.
Asymptotic stability results in Proposition \ref{prop:fixed-point}
still apply upon substituting the vector of winning probabilities
${\bm b}$, which was implicitly depending on fixed parameter $K$, with its average with respect to $K$, i.e.,
$\mathbb{E}_K [{\bm b}(K)]$.

Moreover, one can check whether a given distribution of $K$ can only lead to the natural permutation.
In particular, the critical permutation $1,\ldots,N,N-1$ is not stable provided that
\begin{equation}\label{eq:mixk}
\sum_k p_k \left[2 \left(1-\frac{1}{G 2^{\alpha}}\right)^k -1 - \left(1-\frac{1}{G} - \frac{1}{G 2^{\alpha}}\right)^k\right] < 0
\end{equation}
and one can numerically check whether \equaref{mixk} indeed holds under a given
distribution $\{p_k\}_k$.

\subsubsection{Extension to {\em without-item-repetition}.\\}
The case {\em without-item-repetition} can in principle be handled in the same way as {\em with-item-repetition}, but unfortunately the
exact computation of winning probabilities $b_i$ requires now the enumeration of all
dispositions of $K$ objects from $N$ objects, whose number becomes intractable for large values of $N$,$K$.
Therefore, we propose the following approximate computation of $K_{\min}$ for the case {\em without-item-repetition},
which turns out to be very accurate when compared to exact (computationally feasible) results.
The approximation is based on the following idea: we assume that the top two objects are extracted without repetition, whereas
all other objects can be extracted with repetition. By so doing, we capture the fact that the
two most important objects for our purposes are correctly extracted without repetition,
while all of the others, whose precise identity is not important to determine
winning probabilities $b_N$,$b_{N-1}$, are approximately extracted with repetition
to allow scalable computation.
Let $p^* = 1 - p_N - p_{N-1}$ be the probability to choose a ball within
set $\{1,\ldots,N-2\}$.

Given an extraction of $K$ balls, we have that neither object $N$ nor object $N-1$ wins
with probability $(p^*)^K$.
Then, if we determine $b_{N-1}$, we can easily obtain the complementary $b_N = 1 - (p^*)^K - b_{N-1}$.
Therefore, $b_N \geq b_{N-1}$ is equivalent to $b_{N-1} \leq \frac{1 - (p^*)^K}{2}$.
To compute the winning probability of object $N-1$, we consider
that object $N-1$ wins when: $m$ objects different from $N$,$N-1$ are initially extracted, $0 \leq m \leq K-1$
(and put back in the urn); object $N-1$ is extracted (and not put back in the urn); $N-1-m$ objects different from $N$
are extracted with {\em renormalized} probability $p^*/(1-p_{N-1})$.

It follows:
\begin{eqnarray*}
b_{N-1} \sim \sum_{m = 0}^{K-1} (p^*)^m p_{N-1} \left(\frac{p^*}{1-p_{N-1}}\right)^{K-m-1} =
(p^*)^{K-1}\frac{1-(1-p_{N-1})^K}{(1-p_{N-1})^{K-1}}
\end{eqnarray*}
and we obtain the approximate value of $K_{\min}$:

\begin{equation}
K_{\min} = \min_K \left\{ K: 2 \left( \frac{1-G^{-1}-G^{-1} 2^{-\alpha}}{1-G^{-1}} \right)^{K-1}
\leq \frac{1 - (1-G^{-1}-G^{-1}2^{-\alpha})^K}{1 - (1-G^{-1})^K} \right\}. \label{eq:kminnorep}
\end{equation}

Under the above approximation of $K_{\min}$ for the case {\em without-item-repetition},
all properties listed in Proposition \ref{prop:kmin} still holds, as one can check along the same lines of the proof
reported in Appendix \ref{app:kmin}.
The fundamental difference with respect to {\em with-item-repetition} is the behavior of
$K_{\min}$ as $\alpha$ grows large:
\begin{corollary}\label{coro:bigalfanorep}
For given $N>1$, in the case {\em without-item-repetition}, under approximation \equaref{kminnorep},
$K_{\min} \rightarrow 2$ as $\alpha \rightarrow \infty$.
\end{corollary}
\begin{proof}
The right hand side of \equaref{kminnorep} is larger than one for any $\alpha$.
The left-hand side of \equaref{kminnorep} is equivalent to $ 2\left(1-\frac{1}{(G-1) 2^\alpha}\right)^{K-1}$.
Since $(G-1)2^\alpha = 1 + \sum_{i = 3}^{\infty} (\frac{2}{i})^\alpha$ tends to 1 as $\alpha \rightarrow \infty$,
the left hand side of \equaref{kminnorep} tends to zero as $\alpha \rightarrow \infty$,
hence $K=2$ is enough to make the inequality valid for large $\alpha$.
\end{proof}

\begin{figure}[htb]
\centering
\includegraphics[width=9cm]{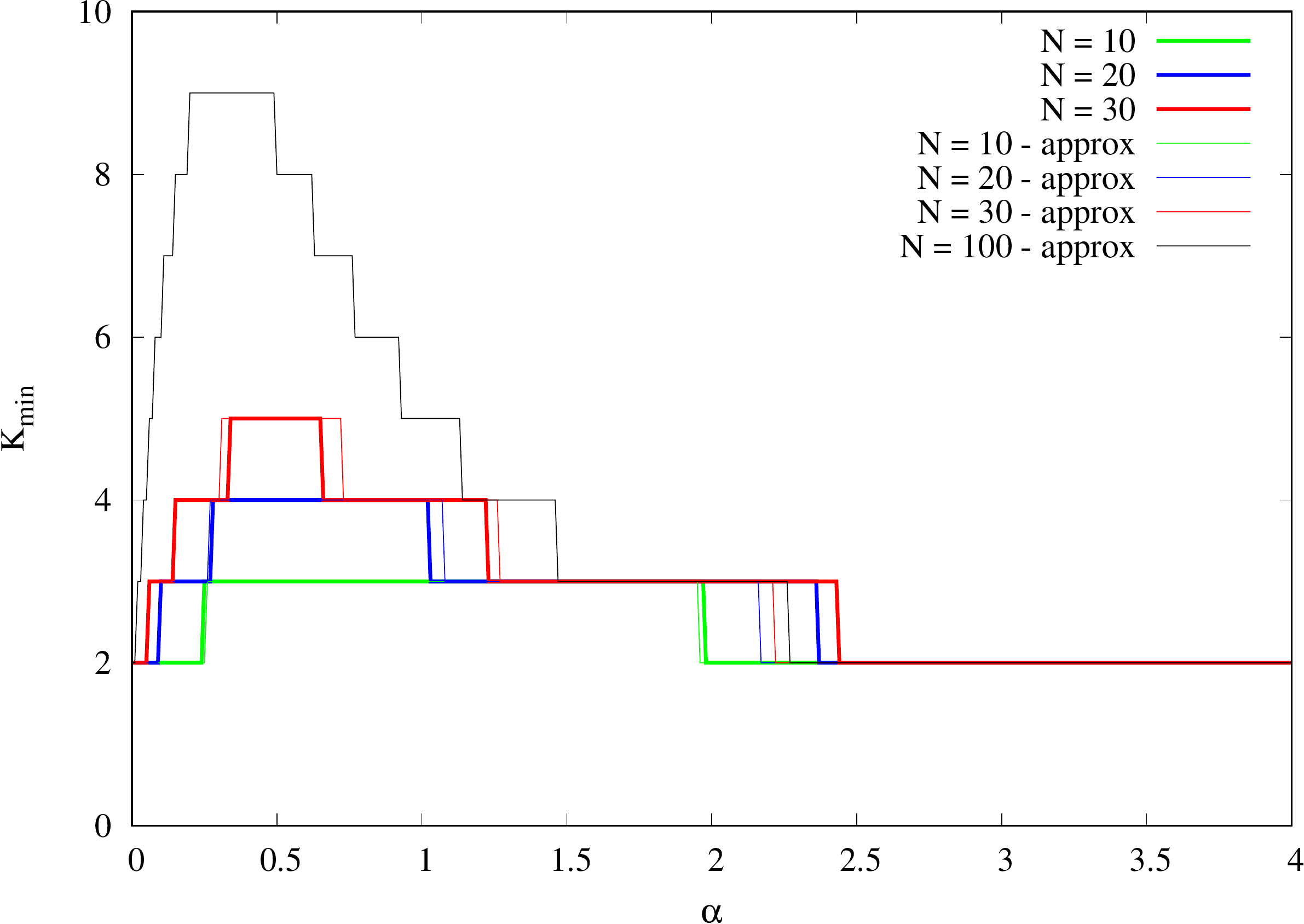}
\caption{Minimum value of $K$ such that the natural permutation is the only stable point,
for different values of $N$, {\em without-item-repetition}. Thin curves refer to approximation \equaref{kminnorep},
whereas thick curves show exact results (when computation is feasible).\label{fig:minkalfanorep}}
\end{figure}

Figure \ref{fig:minkalfanorep} shows $K_{\min}$ as function of $\alpha$, for different
values of $N$, obtained either by the approximate inequality \equaref{kminnorep} (thick curves), or by
exact numerical computation of all stable points, which is feasible only for
small values of $N$,$K$ (we were not able to obtain exact results beyond $N  = 30$).
A number of interesting observations can be drawn from Figure \ref{fig:minkalfanorep}:
i) the approximate formula \equaref{kminnorep}, which can scale to arbitrarily large values of $N,K$, gets very close
to exact values, except for little discrepancies around the points at which $K_{\min}$ varies by 1;
ii) properties in Proposition \ref{prop:kmin} are indeed verified also for {\em without-item-repetition}, and in particular
$K_{\min}$ still diverges for $0 < \alpha \leq 1$; iii) for fixed $N$ and $\alpha$, $K_{\min}$ is smaller under
{\em without-item-repetition} with respect to {\em with-item-repetition}, as one can check by comparison
with Figure \ref{fig:minkalfa}; iv) in contrast to {\em with-item-repetition}, $K_{\min} = 2$ for large $\alpha$, as predicted by Corollary \ref{coro:bigalfanorep}.

\begin{remark}
Results in Figure \ref{fig:minkalfanorep} confirms the existence of a harmful regime of mild popularity bias
($0 < \alpha < 1$) also in the more realistic case {\em without-item-repetition}. To give the reader an intuitive
explanation of this phenomenon, consider the case $K=2$. Denote with $H$ the highest-quality object,
$M$ the second highest-quality object, and $L$ a generic object of lower quality.
The critical permutation in which $M$ becomes more popular than $H$ can be a system stable point
for $N$ sufficiently large and, say, $\alpha$ around 0.5. Indeed, suppose to start
already in this configuration. Then $H$ will struggle to regain popularity versus $M$, because
pre-selections of type $\{M,L\}$ (note that there are many different choices of low-quality item $L$),
which reinforce the popularity of $M$ vs $H$, become jointly more likely than those
of type $\{M,H\}$ or $\{H,L\}$, which would reinforce the popularity of $H$.

Note however that, in the extreme case of $\alpha=0$, any pre-selection of the $N$ objects is equally likely to be generated,
and those in which $H$ wins are more numerous than those in which $M$ wins (indeed,
$H$ wins also in the direct match $\{M,H\}$). Hence for $\alpha$ sufficiently small
we expect that the only stable point will be the natural permutation.

For large $\alpha$, the stability of the critical permutation depends
on the fact that item repetition is allowed or not.
First of all, note that for large $\alpha$
the probability to pre-select an item is concentrated on the most popular items.
In {\em without-item-repetition}, for large $\alpha$ the direct match $\{M,H\}$ will appear more and more frequently, while pre-selections containing also low quality objects will become negligible, so the critical permutation
cannot be stable, and $K_{\min}$ tends to 2.
Indeed, consider an extremely large $\alpha$, and suppose to start from the
critical permutation: after extracting $M$, $M$ can no longer be extracted, and
we have to choose as second object $H$, which wins over $M$.
In {\em with-item-repetition}, the opposite behavior occurs for very large $\alpha$:
now almost all extractions of the first two objects will both be $M$, reinforcing
$M$ itself. Actually for any $K$ there will be an $\alpha$ large enough that
with high probability we extract $M$ consecutively $K$ times, making
the critical permutation stable. Hence {\em with-item-repetition}
$K_{\min} \rightarrow \infty$ as $\alpha \rightarrow \infty$ (corollary
\ref{coro:bigalfarep}).
Of course in a real system that randomizes the list of items proposed to the user
objects are never repeated, so {\em without-item-repetition}, though
more difficult to analyze exactly, is more meaningful.
\end{remark}

\begin{example}
\label{example:application}
As an example of possible application of our results, consider a Video-On-Demand
platform offering $100$ titles in response to queries for a given movie genre.
Suppose that the system shows to users a randomized list of the above titles,
using as popularity metric the current number of views.
If we want the best-quality movie (assuming that this notion exists for movies) to always emerge
from the competition with other movies of the same genre (which occurs when the natural permutation is the only
stable point), the popularity-based amount of randomness introduced in the generation
of the recommendation list (the value of $\alpha$ in our ranking model)
must be carefully controlled. For example, from Figure \ref{fig:minkalfanorep} we see that
values of $\alpha$ around 0.5 are a very bad choice,
unless the quality discrimination power of the users is large ($K \geq 9$).
\end{example}

\section{Analysis of the average quality index}\label{sec:fairness}
The performance of an online cultural market depends on a complex combination
of factors including the way items are internally ranked,
the way in which they are visually presented to the user in response to a given query,
an the user behavior in exploring and selecting among alternatives.
In our simple model, the system has been described by tuple $(N,K,\alpha)$. Among this set,
$\alpha$ can be considered as a parameter tunable by the online platform, and one
can naturally ask whether an optimal value $\alpha^{*}$ can be computed
so as to maximize a given performance metric.

In previous section, we have considered as primary objective the guaranteed emergence
of the natural alignment between popularity and quality.
As a secondary objective, we consider here another important optimization criterion,
which is the average quality of items ultimately chosen by users. Indeed,
it might not be desirable to just maximize such average quality, i.e., to make the top-quality item
monopolize the market. In general, a system might prefer to achieve a desired
level of \lq fairness' among the different alternatives, so that also
mid-quality items have non-negligible chances to be chosen.

Note that in this section we will analyze the average quality index assuming that
the system has a single stable point (the natural alignment).
In Section \ref{sec:multiclass} we will extend the definition of average quality index
to the case in which there are multiple stable points, when we will introduce
the extension of the model to multiple classes of users.

\subsection{Obtaining a desired average quality}\label{subsec:matchquality}

Assuming that $K \geq K_{\min}$, so that the system is guaranteed to achieve the natural alignment, the average quality index $\overline{q}$ takes value in the range $[1,N]$. Since the extreme case $\overline{q} = N$, corresponding
to a system in which only the top-quality item is chosen, is in general undesirable, we assume
that our goal is to achieve a given, intermediate value $1 \leq \tilde{q} \leq N$.

To understand whether a given $\tilde{q}$ is indeed feasible, we need to compute
how metric $\overline{q}$ associated to the natural permutation depends on system parameters $(N,K,\alpha)$, separately
considering the cases {\em with-item-repetition} and {\em without-item-repetition}.

\begin{proposition}[Average quality index under {\em with-item-repetition}.]\label{prop:aveq}
For given $N$ and fixed $K$, the average quality $\overline{q}$ of the natural permutation is an increasing function of $\alpha$.
Possible values of $\overline{q}$ lie in the range $[\overline{q}_{\min},N)$,
where $\overline{q}_{\min}$ is the value attained with $\alpha = 0$:
$$ \overline{q}_{\min} = N - \frac{\sum_{i=1}^{N-1} i^K}{N^K} $$
while $\overline{q} \rightarrow N$ as $\alpha \rightarrow \infty$.

For given $N$ and $\alpha$, the average quality $\overline{q}$ of the natural permutation is an increasing function of $K$.
\end{proposition}
\begin{proof}
See Appendix \ref{app:aveq}.
\end{proof}

Under {\em without-item-repetition}, a formal proof of the monotonicity of $\overline{q}$
with respect to $\alpha$ and $K$ is more difficult, since winning probabilities $b_i$'s lack a simple
closed form expression. However, we have numerically verified that $\overline{q}$ indeed
increases with $\alpha$ and $K$, and results analogous to those stated in Proposition \ref{prop:aveq} holds under
{\em without-item-repetition}. However, the lower extreme of possible values of $\overline{q}$
is different, since here, for $\alpha = 0$, we get:
\begin{equation}\label{eq:aveqwithout}
\overline{q}_{\min} = \sum_{i=K}^N i \cdot  \frac{\binom{i-1}{K-1}}{\binom{N}{K}} =
\frac{\sum_{i=K}^{N} K \binom{i}{K}}{\binom{N}{K}}   = K \frac{\binom{N+1}{K+1}}{\binom{N}{K}}  = K \frac{N+1}{K+1}.
\end{equation}
It is interesting to see how $\overline{q}_{\min}$ in \equaref{aveqwithout} depends on $K$:
with $K=1$ we obtain $\frac{N+1}{2}$, which is the baseline performance
of a system in which quality does not come into play, and any item is equally likely to win the
competition.\footnote{Recall that with $K=1$ any permutation is equally likely
to emerge, but since $b_i = 1/N$ for any item, the specific permutation is not important.}
For $K = N$, the top-quality item always win, and we obtain, as expected, the maximum
possible value $N$.

As immediate consequence of Proposition \ref{prop:aveq}, for given $N$ and $K$,
any desired average quality index $\tilde{q} \in [\overline{q}_{\min}(N,K),N)$ can be obtained
by a unique choice of $\alpha^{*} \in \R^+ \cup {0}$, where we have specified for clarity the
dependency of $\overline{q}_{\min}$ on both $N$ and $K$.

\subsection{Approaching a desired winning probability distribution}

Instead of the average quality index, one could try to obtain a specific
winning probability $\tilde{b}_i$ for each item, assuming that $\{\tilde{b}_i\}_i$ form an increasing sequence with $i$, corresponding to a system operating under the natural permutation.

As an example, we consider a family of desired winning probabilities $\{\tilde{b}_i\}_i$ specified
by a rank-based power-law of exponent $\beta>0$:
\begin{equation}\label{eq:betalaw}
\tilde{b}_i \triangleq \frac{(N-i+1)^{-\beta}}{\sum_{j=1}^N j^{-\beta}} \qquad \text{[desired winning probability]}.
\end{equation}

Here, $\beta$ is a pre-defined exponent reflecting the desired level of fairness:
when $\beta = 0$ we have the extreme case in which all items have the same
winning probability, irrespective of their intrinsic quality; for $\beta \rightarrow \infty$
we approach the other extreme case in which only item $N$ wins.
Correspondingly, we have the desired average quality index
$\tilde{q} \triangleq  \sum_{i=1}^N i \cdot \tilde{b}_i$.

Since winning probabilities $\tilde{b}_i$ cannot be perfectly achieved, we consider as optimal $\alpha^{*}$ the value
of $\alpha$ that minimizes a suitable distance metric between probability distributions $\{b_i\}_i$ and $\{\tilde{b}_i\}_i$.
We will also consider the difference:
\begin{equation}\label{eq:deltaq}
\Delta q \triangleq |\tilde{q}-\overline{q}|
\end{equation}
between the desired average quality index and the actual average quality index.

\subsection{An optimization example}

As an example, we consider an online cultural market containing $N=20$ items where the user discrimination parameter is $K=5$.
Suppose that our goal is to approach the winning probabilities $\{\tilde{b}_i\}$ obtained by using $\beta = 2$ in \equaref{betalaw}, i.e.,
the second-best item has winning probability equal to $1/4$ of the winning probability of the best item.

From Figure \ref{fig:minkalfanorep} we observe that, in the case {\em without-item-repetition},
$K=5$ is enough to guarantee emergence of the natural permutation for {\em any} $\alpha$, so here we do not
need to check whether we indeed have $K > K_{\min}$.\footnote{In general, $\alpha^{*}$ has to be
found in the restricted set of $\alpha$ values for which $K > K_{\min}$.}

\begin{figure}[htb]
\centering
\includegraphics[width=9cm]{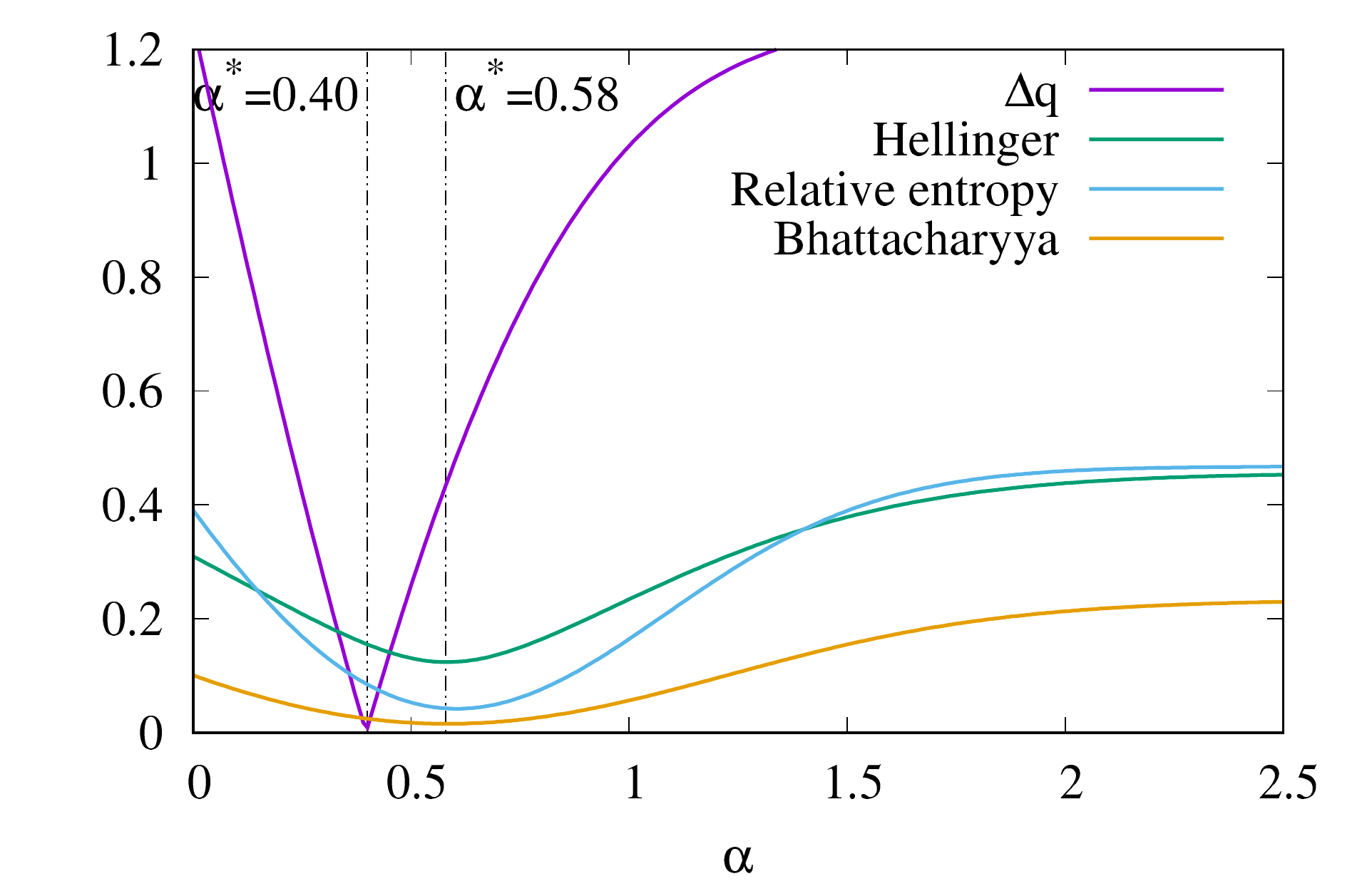}
\caption{Distance between probability distributions $\{b_i\}$ and $\{\tilde{b}_i\}$ as a function of $\alpha$.\label{fig:distance-vs-alpha-without-fm0}}
\end{figure}

Figure \ref{fig:distance-vs-alpha-without-fm0} shows the results of our numerical optimization.
We plot the value of three distance metrics between discrete probability distributions:
\[
\frac{1}{\sqrt{2}} \sqrt{\sum_{i=1}^{N} (\sqrt{b_i} - \sqrt{\tilde{b}_i})^2}
\qquad \text{[Hellinger distance]}
\]
\[
\sum_{i=1}^{N} b_i \log(\frac{b_i}{\tilde{b}_i})
\qquad \text{[Relative entropy]}
\]
\[
-\ln(\sum_{i=1}^{N} \sqrt{b_i \tilde{b}_i})
\qquad \text{[Bhattacharyya distance]}
\]
as functions of $\alpha$. Interestingly, all considered distances between probability distributions $\{b_i\}$ and $\{\tilde{b}_i\}$ reach their
minimum value for $\alpha^{*}$ close to $0.58$. At the same time, there exists a unique value for $\alpha$ ($\alpha^{*}=0.40$)
that nullifies $\Delta q$, as a consequence of Proposition \ref{prop:aveq}.
Note that both values of $\alpha^{*}$ computed above fall in the harmful regime $0 < \alpha \leq 1$: though we can safely
operate in this regime with just $20$ items, doing so with larger values of $N$ could be unfeasible, since
$K_{\min}$ would become too large (see Figure \ref{fig:minkalfanorep}).

\section{Heterogeneous perception of quality}\label{sec:multiclass}
So far we have assumed the existence of an intrinsic quality metric of each item, such that
all users would rank the $N$ items in the same order (provided that they had the time to
independently examine all of them). Since quality is highly subjective, and
its assessment may vary among users, we now extend the model to the case of a (finite) number $C$ of user classes:
the user arriving at time $n$ is assumed to belong to class $1 \leq c \leq C$ with probability $f_c$, independently
of other users. Of course we have the constraint $\sum_{c=1}^{C} f_c = 1$.

Users belonging to the same class are assumed to perceive items' quality in such a way that they
would all rank the items in the same order  (provided that they had the time to
independently examine all of them). Therefore, user class $c$ is characterized by a unique
(fixed) vector ${\bm v}_c$, listing the items' id's in increasing order of their quality, as perceived by users belonging to that class. Note that in the base setting there is a single class ($C = 1$),
with ${\bm v}_1 = [1,2,\ldots,N]$.
Our model is quite general, since we do not impose any restriction on vectors ${\bm v}_c$, that could
differ completely from one class to another. For example, users belonging to a class may perceive an item as high-quality,
while users belonging to another user may perceive the same item as low-quality.

It turns out that the multi-class version of the system can be analyzed
essentially in the same way as the single-class instance, by substituting
the winning probability of each item by a weighted average (where weights correspond to probabilities $\{f_c\}$)
of the winning probabilities computed for the same item according to the different classes.
Therefore, we can introduce a generalized mapping function $B(.)$ as done in Section \ref{sec:popularity-pre-selection}
and easily extend to the multi-class case the Definition \ref{def:stablepoint} of system stable point.

The numerical computation of system stable points can thus be performed in a similar way as for the single-class setting.
Algorithm \ref{alg:search-multiclass} is a simple and quite straightforward generalization of Algorithm \ref{alg:search} to the
case of multiple user classes.

\begin{algorithm}[thb]
\begin{algorithmic}[1]
\Require $N \in \N^+$, $K \in \{1,\ldots,N\}$, $\alpha \in \R^+$
\State Choose random positive values \mbox{$\{\tilde{w}_i\}_{i=1}^{N}: \sum_{i=1}^{N} \tilde{w}_i = 1$} \Comment{Initialization of $w_i$'s}
\State $fixedpoint \Leftarrow \FALSE$
\While{($fixedpoint = \FALSE$)}
\State $r_i \Leftarrow$ index of $i$ in sorted $\tilde{\bm w}$ , $r_i \in \{0,\ldots,N-1\}$
\State $p_i \Leftarrow (N-r_i)^{-\alpha}, \forall i$ \Comment popularity of $i$
\State normalize $p_i$'s such that $\sum_{i=1}^{N} p_i = 1$
\State $\bm b \Leftarrow \bm 0$
\For{c=1 \textbf{to} C}
\State $p_{i,c} \Leftarrow p_{{\bm v}_c[i]}$, $i \in \{1,\ldots,N\}$ \Comment popularity of objects, ordered by quality, according to class $c$
\State compute $b_{i,c} = \PP[\text{object} \,\,i\,\, \text{wins}]$, $i \in \{1,\ldots,N\}$  \Comment using either \equaref{birep} or \equaref{bidis}, and $p_{i,c}$ in place of $p_i$
\State $b_{{\bm v}_c[i]} \Leftarrow b_{{\bm v}_c[i]} + f_c b_{i,c}$, $i \in \{1,\ldots,N\}$ \Comment contribution of class $c$ to winning probability of object $i$
\EndFor
\If{$\bf b = \tilde{\bm w}$}
\State	$fixedpoint \Leftarrow \TRUE$
\Else
\State $\tilde{\bm w}\Leftarrow \bm b$
\EndIf
\EndWhile
\If{$b_i \neq b_j, \forall i \neq j$}
\State $\bm b$ is a stable point
\EndIf
\end{algorithmic}
\caption{Randomized computation of equilibria with multiple users classes}
\label{alg:search-multiclass}
\end{algorithm}

Unfortunately, instead, the analytical results on $K_{\min}$ derived in Section \ref{subsec:kmin}
no longer apply to the multi-class case, since they are crucially based on the assumption
that there exists a unique quality ranking for the items.
Given the wide variety of multi-class scenarios that can be considered (notice that, beyond
the number $C$ of classes, detailed characteristics of each class as specified
by vector ${\bm v}_c$ can affect the transient and asymptotic system behavior),
we have limited ourselves to a preliminary investigations of just a few scenarios.

As an example, we have considered the case of $N = 10$ items,
with at most three user classes whose characteristics are summarized in Table \ref{tab:3classes}:
for each class $c$, $c = 1,2,3$, the id's of the items, ordered by quality as perceived by users of class $c$ (i.e., vector ${\bm v}_c$),
are listed from the lowest-quality (left-most item) to the highest quality (right-most item).
Note that vectors ${\bm v}_c$ where chosen simply at random from the space of 10! permutations.

\begin{center}
\begin{table}[htb]\centering
\begin{tabular}{l||c|c|c|c|c|c|c|c|c|c|}
 (class 1) ${\bm v}_1$: & 5 & 7 & 4 & 9 & 8 & 1 & 2 & 6 & 3 & 0 \\ \hline
 (class 2) ${\bm v}_2$: & 3 & 1 & 2 & 5 & 9 & 7 & 4 & 8 & 0 & 6 \\ \hline
 (class 3) ${\bm v}_3$: & 7 & 9 & 0 & 4 & 2 & 3 & 8 & 1 & 5 & 6 \\
\end{tabular}
\caption{Items ordered by increasing quality, according to 3 users classes}\label{tab:3classes}
\end{table}
\end{center}

\begin{figure}[htb]
\centering
\includegraphics[width=14cm]{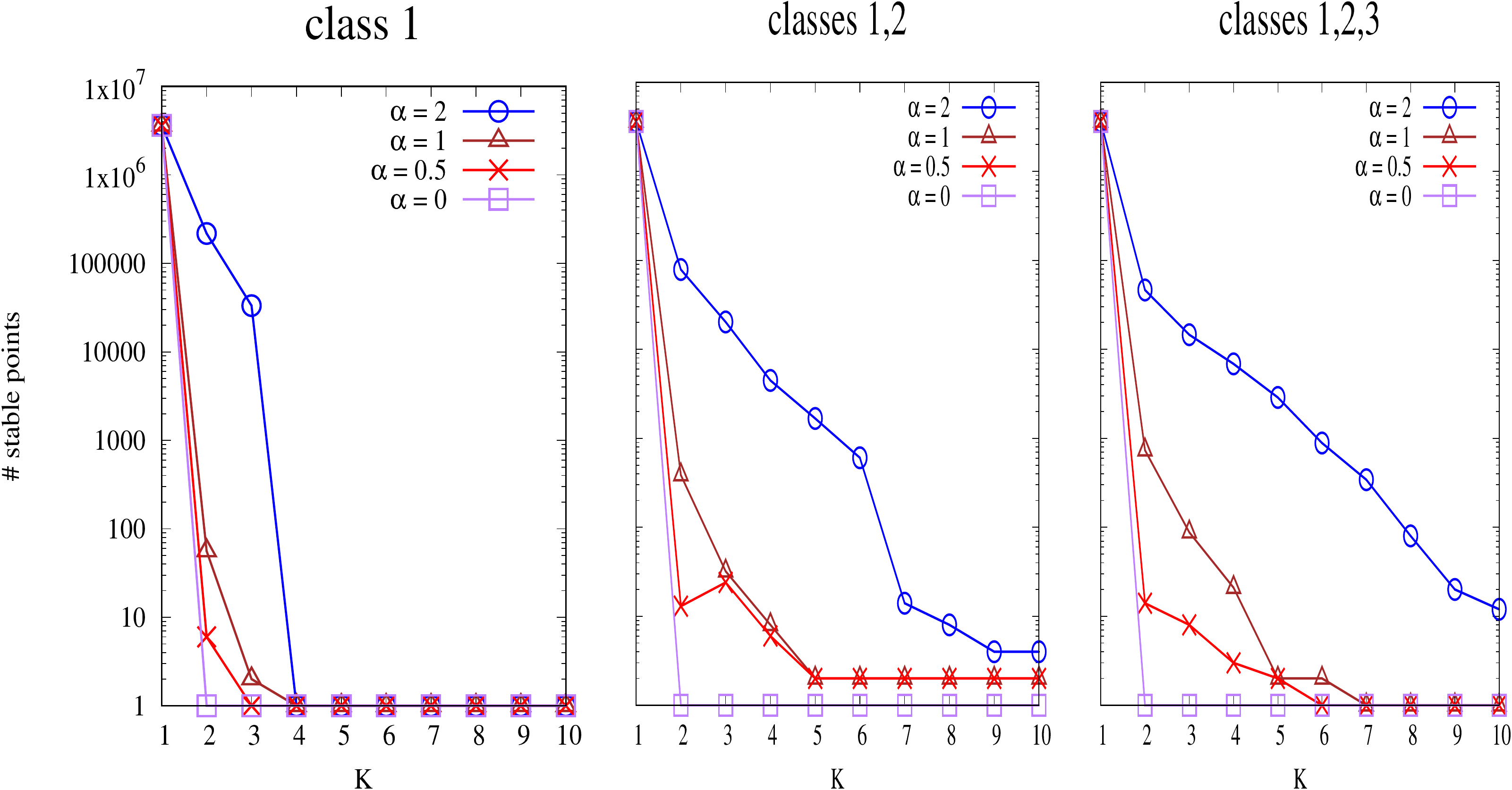}
\caption{Number of system stable points as function of $K$, for different values of $\alpha$, in the case of $N=10$, {\em with-item-repetition}: with only class 1 (left plot); with classes 1,2 (middle plot); with classes 1,2,3 (right plot).\label{fig:classwithrep}}
\end{figure}

Figure \ref{fig:classwithrep} reports, on a log vertical scale, the number of system stable points as function of $K$,
for 4 different values of $\alpha = 0, 0.5, 1, 2$, comparing 3 systems with $N=10$, {\em with-item-repetition}:
a system with $C = 1$ (only class 1, left plot); a system with $C = 2$ (classes 1 and 2, middle plot);
a system with $C = 3$ (classes 1,2, and 3, right plot).
In all scenarios we assume that classes are equally represented (i.e., $\forall c, f_c=\frac{1}{C}$).

Note that the system with $C=1$, equivalent to the single-class case, confirms
results in Figure \ref{fig:minkalfa}, according to which $K_{\min} = 2$ for $\alpha = 0$,
$K_{\min} = 3$ for $\alpha = 0.5$, $K_{\min} = 4$ for $\alpha = 1$ or $\alpha = 2$.
When $K < K_{\min}$, it is interesting to observe how the number of stable points (equal to 10! for $K=1$) is drastically
reduced by increasing $K$.
A similar strong impact of $K$ on the number of stable points is observed also
in the other two cases with multiple classes, but the decay is slower, suggesting that
systems with wildly different user classes are much more difficult to stabilize.
Moreover, while the system with $C=3$ can actually reach a unique configuration for $K$ sufficiently large (except for $\alpha=2$),
the system with $C=2$ always leads to multiple stable points, except for $\alpha = 0$, for which
a unique configuration is stable provided that $K \geq 2$ (this special behavior for $\alpha = 0$ holds in general).

\begin{figure}[htb]
\centering
\includegraphics[width=14cm]{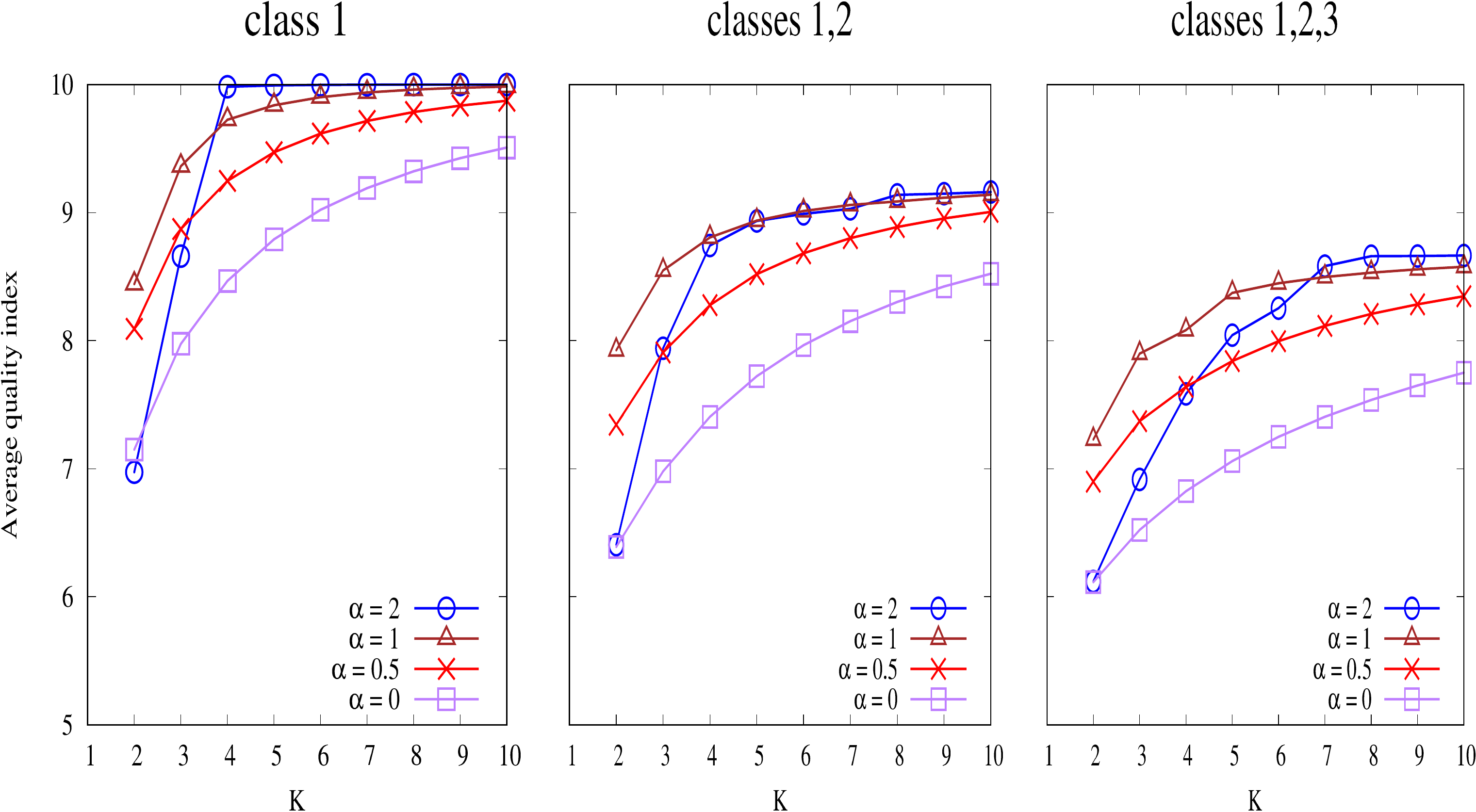}
\caption{Average overall quality index $\overline{Q}$ as function of $K$, for different values of $\alpha$, in the case
of $N=10$, {\em with-item-repetition}: with only class 1 (left plot); with classes 1,2 (middle plot); with classes 1,2,3 (right plot).\label{fig:qualitywithrep}}
\end{figure}

Interestingly, for $C=2$, $\alpha=0.5$, the number of
stable points is even non-monotonic with $K$. This anomaly prompted us to investigate also
the behavior of the average quality index, Figure \ref{fig:qualitywithrep}.
In the presence of multiple stable point, the overall average quality index $\overline{Q}$ is computed as the
weighted sum of average quality indexes of individual stable points, weighted by their attractiveness:
$$\overline{Q}=\sum_{f} a(f)\,\overline{q}(f)$$
Recall from Section \ref{sec:popularity-pre-selection} that attractiveness is defined as the number of
permutations falling into a given fixed point by recursive applications of function $B(.)$. We use it as a proxy of the
likelihood that the system converges to a given equilibrium point, which is unfortunately hard to
characterize since it depends on initial weights ${\bm w}[0]$.

Interestingly, $\overline{Q}$ always increases with $K$, suggesting, in line with intuition, that
stronger quality-discrimination of users leads to better average quality of selected items,
also in the multi-class case. However, as we increase the number of classes, achievable values
of $\overline{Q}$ tend to decrease, as consequence of the fact that a certain stable point (unique or not)
inevitably cannot make happy users belonging to many (very different) classes.

At last, we performed also simulations of the above multi-class scenario, to derive performance metrics
that cannot be easily derived analytically, not even for the single-class case.
One such metric is the distribution of the smallest time for the system to reach one of its stable points.
We assume that the system starts from the initial condition in which popularity weights are all equal to 1, i.e., $\forall i, w_i[0]=1$.
We simulated $10^6$ run for each scenario, as we vary the number $C$ of classes and parameter $\alpha$.

Figure \ref{fig:multi-class-simulation-median} shows the median of the distribution of the smallest time to reach any stable point,
as function of parameter $\alpha$, for $N=10$, $K=5$, {\em without-item-repetition}.
We observe that, as a general trend, the stronger the popularity bias represented by $\alpha$, the longer it takes to the system to reach one of its
stable points. One exception is in the case $C=3$ (note that the median diminishes passing from $\alpha = 0.5$ to $\alpha=0.6$): this can be
explained by the fact that, in the interval $\alpha \in [0.6,0.8]$, the system has two stable points instead
of 1 (intuitively, the higher the number of system stable points the lower the time to reach any of them).

Figure \ref{fig:time-simulation} shows, on a log-log scale, the evolution over time of normalized weights $\tilde{w}_i[n]$ in a particular
simulation run obtained for one of the scenarios considered in Figure \ref{fig:multi-class-simulation-median}, i.e., $C=3$, $\alpha = 0.5$.
We observe that, after a turbulent initial phase characterized by large
randomness affecting mid-popularity objects,
trajectories converge to their asymptotic constant value, becoming almost
flat after about $10^6$ user interactions.

\sidebyside{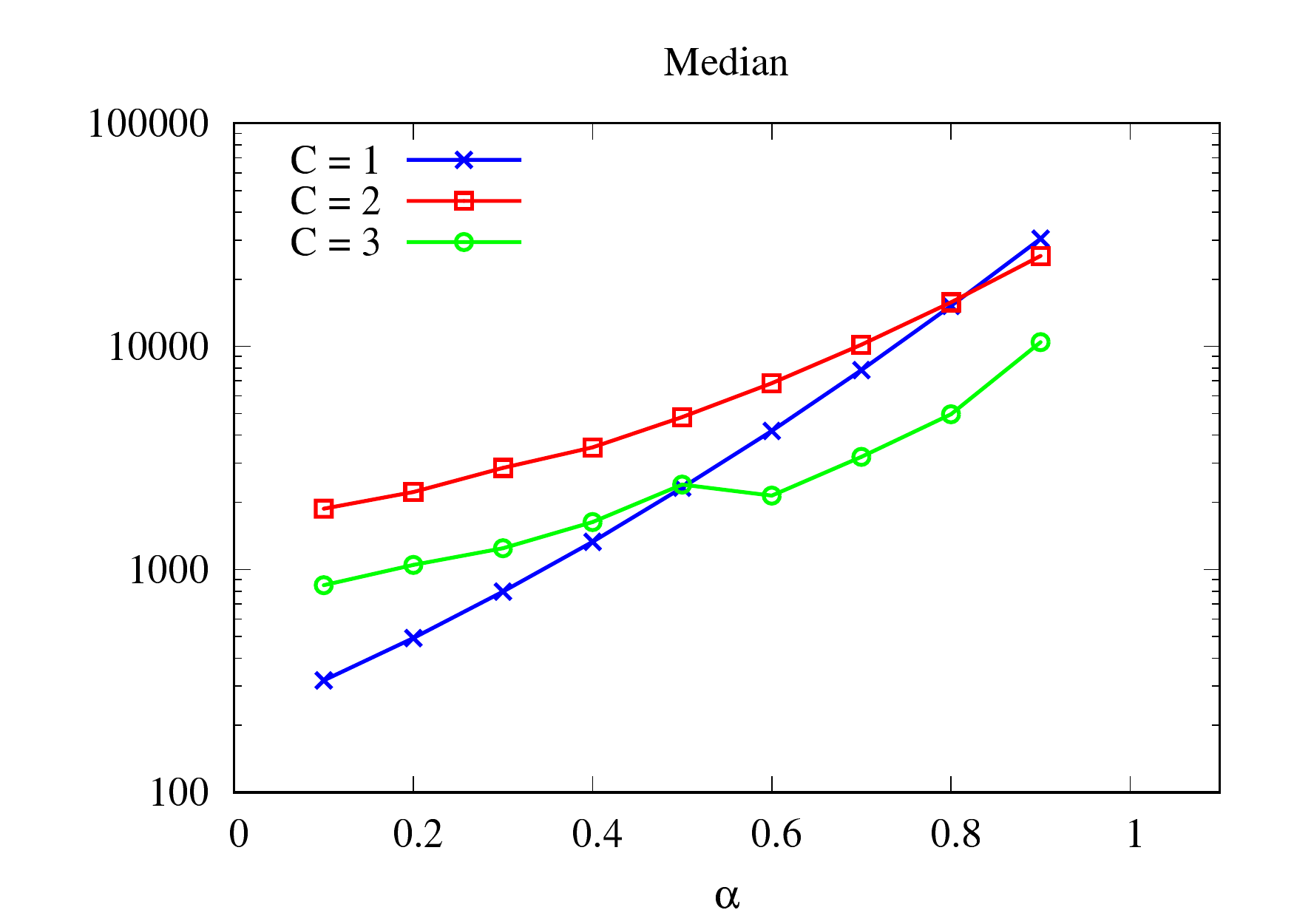}{Median of the minimum time to reach a system stable point as function of $\alpha$,
for different number of classes.}{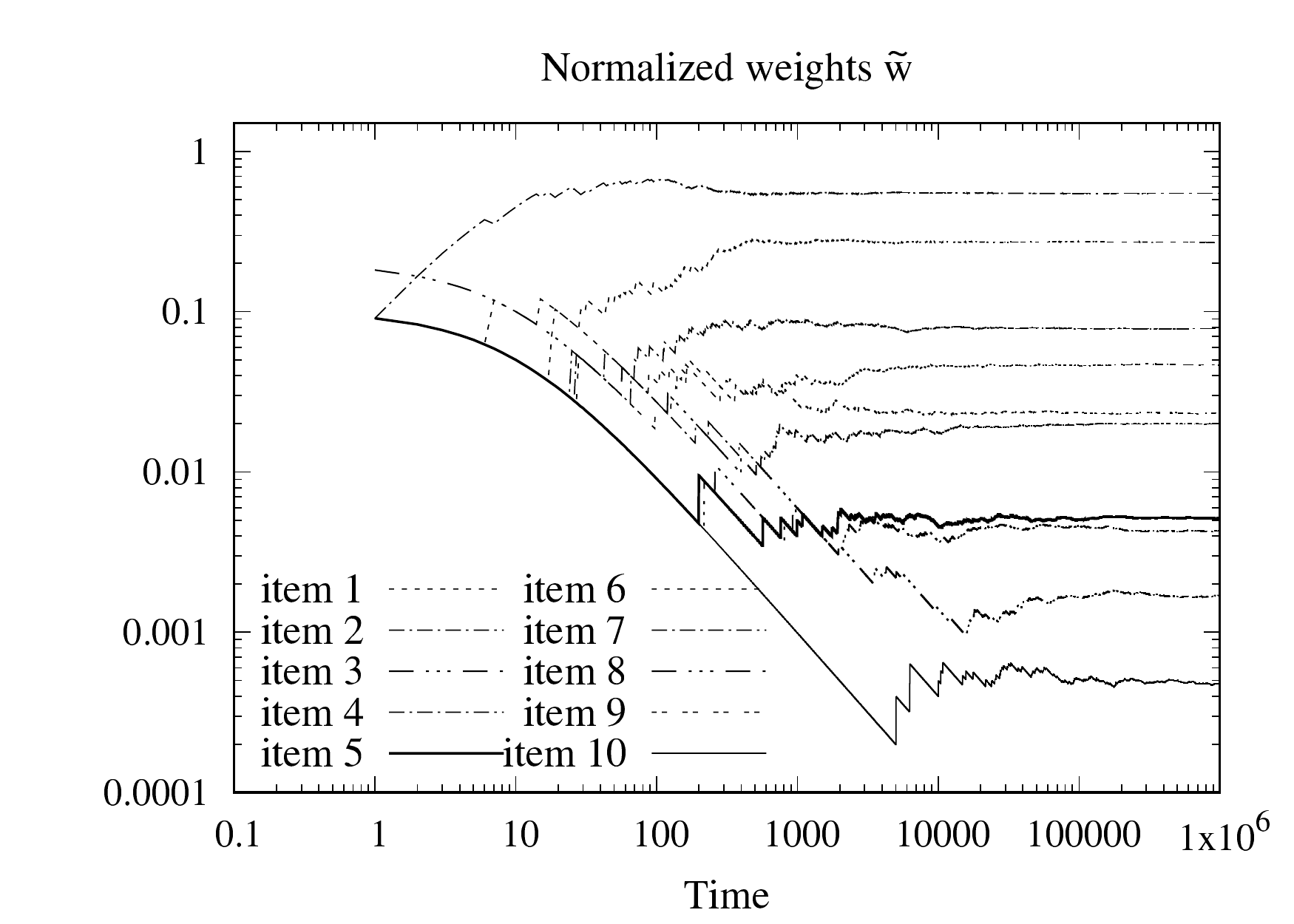}{Evolution of normalized weights $\tilde{w}_i[n]$ as function of time, in the case $C=3$, $K=5$, $\alpha=0.5$.}

\section{Related work}
\label{sec:related}
Our work studies the consequences of using popularity as proxy for quality.
There has been work supporting this view.
In \cite{stoddard15}, for example, authors consider two different social news aggregators, Reddit and Hacker News.
They define quality as the number of votes an article would have received if shown, in a bias-free way,
to an equal number of users. Using a Poisson regression method they find that
popularity on Reddit and Hacker News is a relatively strong reflection of quality.
It is therefore to be expected that, especially in condition of limited attention,
one makes choices using popularity-based heuristics.
There is a vast literature proving that popularity is indeed an important factor that influences choices:
Cai et al., e.g., studied the choices from a restaurant menu and showed that the demand for the most popular
dishes increased when this information was made explicitly available ~\cite{cai_observational_2009}. Salganick and collaborators conducted a randomized experiment based on an artificial market for music downloads
and proved a marked difference in people choices when they are exposed to those of others~\cite{salganik_experimental_2006}; such differences can ultimately result in popularity rankings (for the same set objects) that differ greatly from experiment to experiment and from the ranking that one would observe if choices where made in isolation and therefore independently. Such idiosyncrasies can be attributed to the noisy popularity bias induced by the somewhat random choices made by the first users to perform their choices, and they are the reason that make success hard to predict in cultural markets~\cite{watts_everything_2011}.
We can easily appreciate the powerful effect of noisy initial fluctuations in popularity ranking in modelling cultural market, e.g., in~\cite{Weng2012CompetitionAM} items equally appreciated by generic users can reach very different levels of popularity due to popularity biases. Other studies have observed empirically the somewhat pernicious effect of popularity bias~\cite{lorenz}, or more in general of social influence, in our choices~\cite{muchnik_social_2013}. A related, but distinct, issue is that when popularity informs the recommendation of a recommending systems it can over concentrate the attention of users on the most popular items at the expense of equally valid but poorly recognized others (see e.g.,~\cite{Abdollahpouri2019}). Indeed, even when the recommender system is well aligned with the preference of the majority, it can still happen it produces ``unfairly'' poor recommendation for those whose preferences are misaligned with those of the majority of users~\cite{Abdollahpouri2019etAl}. Finally it is probably interesting to notice that also in situations where a reasonably uncontroversial definition of quality can be assumed (where, e.g. it can be linked to measurable outcomes or performances) popularity bias can provoke misalignments between quality and popularity rankings~\cite{barabasi_formula_2018}.
In~\cite{engstrom_demand_2018}, a study about apps downloads from Google Play, the authors have
shown that consumers are more sensible to others revealed rather than stated preferences. A similar result is reported for hotels choices in ~\cite{viglia-please-2014}.

Many different, often overlapping, reasons lay behind our tendency to inform our choices by popularity.
First, people's behavior carries information and there are
circumstances in which it is a perfectly rational and effective strategy to observe and copy, even though this may lead to information
cascades, as demonstrated in the classical works~\cite{banerjee_simple_1992, bikhchandani_theory_1992, welch_sequential_1992}.
Second, adopting popular choices can derive from social
influence ~\cite{bakshy_everyones_2011,phd_connected_2009,krumme_quantifying_2012}, which, in turn, can produce undesirable outcomes such as herding effects,
that have been well documented in many areas of human behavior~\cite{raafat_herding_2009}.  Third, the tendency of
choosing the most popular items may reflect a cognitive \lq\lq bias" or heuristics. In cognitive sciences the notion of
\lq\lq recognition" heuristics has been widely studied. In the words of Goldstone and Gingerenzer~\cite{gigerenzer_simple_2000}:
\lq\lq Consider the task of inferring which of two objects has a higher value on some criterion (e.g., which is faster, higher, stronger).
The recognition heuristic for such tasks is simply stated:
If one of two objects is recognized and the other is not, then infer that the recognized object has the higher value".

While it can be cost and time effective to use popularity as a proxy for quality and value, this does not come without a price.
A side effect of popularity driven dynamics is that it implies a positive feedback loop by which items
that are already popular tend to become even more popular. The net effect of such dynamics is that, even
when it produces popularity rankings that are aligned with the quality of items, it concentrates the collective
attention towards few items at the top. Indeed, since the seminal work of Simon~\cite{simon_class_1955}, the rich-get-richer phenomenon has been quantitatively translated in the principle of linear growth and claimed to explain power law distribution of quantities that can be
regarded as popularity ~\cite{rigney_matthew_2010}. Most recently the principle has found new life in network science, where it is known as preferential attachment~\cite{barabasi_emergence_1999} and has been used to justify the widespread occurrence of power law degree
distributions in network abstraction of real-world systems.

When this effect is brought to the extreme and all popularity is concentrated in only few items at the top of the ranking,
then, naturally, diversity in the system disappears. This danger has been noted in several contexts, including that of recommender systems.
In \cite{hu17} authors deal with the problem of data scarcity affecting items in the tail of the popularity
distribution, and develop techniques to improve their estimated quality and robustness to shilling attacks.
In \cite{abd17} authors propose a regularization-based framework to enhance the long-tail
coverage of recommendation lists in a learning-to-rank algorithm, in order to achieve a desired trade-off
 between accuracy and coverage.
\cite{abd20} further explores the fairness problem in recommendation
by considering how different user groups are affected by algorithmic popularity bias.
The same issues of popularity concentration towards few top-ranking items has been considered in studies
about search engines. These presumably use popularity as a signal of relevance. Also they direct attention towards already popular sites via
the (popularity-based) ranking they produce~\cite{cho_impact_2004}. Also in this case there is an implicit risk of a positive feedback loop, although
this tendency is dampened by the diversity of user interests~\cite{fortunato_topical_2006}.

In the recommendation systems community, it is well known that popularity bias  and feedback loop in a long run operation have important consequences on the performance of the system itself. This has been studied - among the others - by Chaney et al.~\cite{chaney_2018}, Jiang et al.~\cite{jiang_2019}, D’Amour et al.~\cite{damour_2020}, and Mansoury et al.~\cite{mansoury_2020}.

On the empirical side, a systematic study of the interplay between quality and popularity is problematic for several reasons.
The most prominent is that the notion of quality, especially where cultural markets are concerned, is to some extent
subjective and therefore difficult to operationalize. A number of studies have nevertheless confirmed the finding of
the music lab experiment ~\cite{salganik_leading_2008,salganik_web-based_2009,muchnik_social_2013,van_de_rijt_field_2014}.
Motivated by ~\cite{salganik_leading_2008}, a number of theoretical papers have tried to avoid the empirical difficulties by postulating
some mechanism of choice that takes popularity into account. Krumme et al proposed an agent based model that closely reproduces
the findings of the music lab experiment ~\cite{krumme_quantifying_2012}.
Van Hentenryck and collaborators considered a model of market in which customers can try products before committing to buying.
They introduce a policy which successfully recovers quality ranking but asymptotically leads to a monopoly of the top-quality item ~\cite{hentenryck_aligning_2016}.

Popularity bias is not easily eliminated, even when one is aware of it. In \cite{marlin09} authors investigate the non-random missing data problem (NRMD), e.g., the fact that users are more
likely to supply ratings for items that they do like, and less likely to supply ratings for items that they do not like.
Incorrect assumptions about missing data have been found to lead to biased parameter estimation
in collaborative ranking.
In \cite{mena20} authors focus on the impact of false-positives, i.e., suggestions that are disliked by users,
discovering a surprising degree of disagreement with true positives, actually penalizing
the most popular items.
In \cite{should18} authors formally investigate  how popularity bias is affected by
random variables such as item relevance, item discovery by users, and the decision by users to
interact with discovered items.

\section{Conclusions and future work}\label{sec:discussion}
In this paper we have shown that a small effort on behalf of users to discriminate quality out of popularity
can, in the long run, straighten out systems that, by themselves, could drift to undesirable
configurations containing severe misalignments, especially among the top-quality items.

Readers familiar with load balancing strategies might recognize an analogy
between our findings and the so called \lq\lq power of two choices" paradigm \cite{Mitzenmacher},
which has been applied also to other computer science problems,
such as hashing and shared memory machines \cite{power}.
For example, in the classical balls-and-bins model, suppose that
$n$ balls are thrown into $n$ bins. A logarithmic reduction in the maximum number of balls
in any bin is obtained when, instead of throwing each ball uniformly at random, we
select two bins uniformly at random, and put the ball in the least loaded one \cite{azar99}.
Note that in our case the goal is not to obtain a balanced allocation of balls among the bins,
but an allocation satisfying a desired ranking (the one associated to intrinsic quality).
Although our problem is different, the discovered phenomenon
is similar: a minimum, local effort performed during the addition
of an individual ball can, in the long run, rectify the distortions produced by a
purely random choice, guaranteeing to achieve the desired configuration.
In particular, in analogy to the \lq\lq power of two choices", we discovered that $K_{\min} = 2$
is enough to guarantee convergence to the desired configuration in any system run (Section
\ref{sec:random-pre-selection}), assuming that items are selected uniformly at random.

In contrast to classic results for the balls-and-bins model, we have considered also the case in which
the candidate bins to receive a new ball are not selected
uniformly at random, but according to their current load (ranking model of exponent $\alpha$).
Quite surprisingly, we have found that in this case $K=2$ is not always enough to
guarantee convergence to the desired ranking. In particular, for $0 < \alpha < 1$,
$K_{\min}$ is even unbounded as $N$ grows large, while for $\alpha > 1$
we recover the effect that a bounded, small $K \ll N$ is enough (Section
\ref{sec:popularity-pre-selection}), though $K_{\min} = 2$ only for large $\alpha$.
Unfortunately, the regime $\alpha > 1$ might not allow us to achieve the desired level of fairness among items
(Section \ref{subsec:matchquality}).

Our analysis is not without limitations, and several directions could be pursued to
generalize it and make it closer to real systems. For example, we have assumed
that users can always find the best-quality items among a subset of $K$:
one could incorporate also a possibly imperfect outcome of the user evaluation process.
Moreover, the $K$ inspected items might not always be on the top of the list
produced by the system (see example in Section \ref{subsec:toyexample}), and one could
try to combine the effect of users focusing their attention on random items of the list.
Another important point is that modern platforms attempt to automatically estimate item quality
in a way similar to what real users would do, e.g., by analyzing reviews left by other users
(e.g., by sentiment analysis),
and combine such estimates with popularity metrics in building their recommendation lists, de facto boosting the quality discrimination performed by users alone.
Indeed it would be interesting to develop models in which popularity, quality and randomness are jointly combined to determine the subset of items inspected by users.

At last, we acknowledge that real systems do not construct
recommendation lists based on our idealized ranking model (power-law with exponent $\alpha$):
different laws (other than power-law) could be considered, as well
as computationally simpler strategies to present to users randomized lists so as to favor
diversity and allow serendipity. For example, some platforms (e.g., Amazon) seem to just
perturb the list produced by their ranking algorithms by randomly
moving up or down a few items (occasionally adding also some \lq intruders'), so that
each non-repeated query produces a slightly different items view.
It would be interesting to study whether such simpler ways to randomize lists
could be mapped onto our ranking model with properly chosen $\alpha$.
Another interesting direction would be to consider dynamic systems \cite{iniguez2021universal} in which new items
are progressively inserted into the catalog, and study strategies to let them quickly emerge
over older (worse) ones.

Finally, our preliminary investigation in Section \ref{sec:multiclass} suggests that the system behavior
in the case of multiple user classes is much richer and more complex than the single class case. Future work should expand our understanding of the multi-class scenario, and, more in general,
the impact of heterogeneous perception of quality.

\section*{Acknowledgements}

A. Flammini acknowledges support from DARPA award HR001121C0169.

\appendix

\section{Proof of Proposition \ref{prop:increasing}}
\label{app:increasing}

\begin{proof}
Let $r_{\min} \triangleq \min_{i} \frac{b_{i+1}}{b_i} > 1$ be the minimum ratio
between two consecutive (positive) winning probabilities.
For example, in the case of winning probabilities \equaref{birandom}
we have, for $K \leq i < N$,
$$ \frac{b_{i+1}}{b_i} = \frac{i}{i-K+1} $$
hence $r_{\min} = \frac{N-1}{N-K} > 1$.
Moreover, let $b_{\min} \triangleq \min_i\{b_i\}$.

Next, we recall the following classic bounds for the tails of the binomial distribution $\mathrm{Bin}(n,p)$ (see
e.g. Lemma 1.1 p. 16 in \cite{penrose2003random}).
Let $\mu:= n p$, and
\begin{equation}\label{eq:Hx}
H(a):=1-a+a\log a,\quad a\in\R_+.
\end{equation}
For any
$0<k<n$, we have:\\
\noindent if $k\geq\mu$, then
\begin{equation}\label{penroseright}
P(\mathrm{Bin}(n,p)\geq k)\leq\exp\left(-\mu H\left(\frac{k}{\mu}\right)\right);
\end{equation}
if $k\leq\mu$, then
\begin{equation}\label{penroseleft}
P(\mathrm{Bin}(n,p)\leq k)\leq\exp\left(-\mu H\left(\frac{k}{\mu}\right)\right).
\end{equation}

We identify a suitable threshold $t_i$ in between each pair of consecutive
winning probabilities $(b_i,b_{i+1})$. In particular, we set $t_i$ as the
geometric mean of $b_i$ and $b_{i+1}$:
$$ t_i = \sqrt{b_i b_{i+1}}, \quad K \leq i < N $$
and we also set $t_{K-1} = 0$, $t_N = 1$.
We denote by $C^i_n = \{t_{i-1} < \frac{w_i[n]}{n} < t_{i}\}$ the event that, at round $n$,
the empirical frequency of the number of times object $i$ wins is
comprised in the interval $(t_{i-1},t_i)$. Observe that we can safely disregard
initial weights ${\bm w}[0]$, which provide vanishing contribution to empirical
frequencies $\frac{w_i[n]}{n}$.

Note that the joint occurrence
of all events $C^i_n$, $\forall i$, is a sufficient but not necessary condition
for the occurrence of $E_n$, which is the event that weights are correctly ordered.
Therefore, denoting by $\overline{E}$ the complement of a generic event $E$, we have:
$$ E_n \supseteq (\cap_{i=K}^N C^i_n) \Rightarrow \overline{E}_n \subseteq (\cup_{i=K}^N \overline{C}^i_n) $$
By the union bound,
$$ \PP(\overline{E}_n) \leq \sum_{i=K}^N \PP(\overline{C}^i_n).$$
Using the above deviation bounds of the binomial distribution, we have, for $K < i < N$:
\begin{multline}
\PP(\overline{C}^i_n) = \PP(\mathrm{Bin}(n,b_i) \geq n t_i) + \PP(\mathrm{Bin}(n,b_i) \leq n t_{i-1})
\leq \exp\left(-n b_i H(t_i/b_i)\right) + \exp\left(-n b_i H(t_{i-1}/b_i)\right) = \\
\exp\left(-n b_i H(\sqrt{b_{i+1}/b_i}\right) + \exp\left(-n b_i H(\sqrt{b_{i-1}/b_i}\right)
\leq \exp\left(-n b_i H(\sqrt{r_{\min}}\right) + \exp\left(-n b_i H(\sqrt{1/r_{\min}}\right).
\end{multline}
For the special case $i=K$ ($i=N$), we have to consider only the right (left) tail,
obtaining:
$$  \PP(\overline{C}^K_n) = \PP(\mathrm{Bin}(n,b_K) \geq n t_K) \leq \exp\left(-n b_K H(\sqrt{r_{\min}}\right) $$
$$  \PP(\overline{C}^N_n) = \PP(\mathrm{Bin}(n,b_K) \leq n t_{N-1}) \leq \exp\left(-n b_N H(\sqrt{1/r_{\min}}\right).$$
In conclusion, since $b_i \geq b_{\min}$, we have
\begin{eqnarray*}
\PP(\overline{E}_n) \leq (N-1)\Big[ \exp\left(-n b_{\min} H(\sqrt{r_{\min}}\right) +
\exp\left(-n b_{\min} H(\sqrt{1/r_{\min}}\right) \Big]
\end{eqnarray*}
meaning that the probability of event $\overline{E}_n$ decreases exponentially to zero as $n$ increases.
Since \mbox{$\sum_n \PP(\overline{E}_n) < \infty$}, the assert follows from the Borel-Cantelli lemma.
\end{proof}

\section{Proof of Proposition \ref{prop:fixed-point}}
\label{app:fixed-point}

\begin{proof}[Sketch of proof]
We present a simplified proof based on the analysis of the trajectory of mean
(normalized) weights $\bar{\bm w} = \E[\tilde{\bm w}]$. As it usually occurs in
P\'{o}lya urn models, the actual system exhibits wild random fluctuations
at the beginning of the process, but the impact of stochasticity diminishes over time
due to the accumulation of balls in the urns, which makes the randomness
related to the addition of individual balls less and less important, so that the
system behavior becomes more and more concentrated around mean values,
which follow a deterministic trajectory on which we restrict our attention here.

Possible values of normalized weights $\tilde{\bm w}$ lie in the simplex
${\mathcal S} = \{{\bm w} \in \R^N: w_1+\ldots+w_N = 1, w_i \geq 0, i = 1,\ldots,N\}$,
which can be partitioned into $M = N!$ subregions $\{{\mathcal S}_i\}_1^M$ corresponding
to the $N!$ possible permutations of indexes $\{1,\ldots,N\}$
associated to sorted values of $\{\tilde{w}_i\}_i$.
It is immediate to verify that subregions ${\mathcal S}_i$ are convex. At the beginning of round $n$ (i.e., after $n-1$ balls have
been added, starting from $n_0$ initial balls in the system), normalized weights $\tilde{\bm w}[n-1]$
lie in a subregion of ${\mathcal S}$ associated to the vector of winning
probabilities ${\bm b}[n] = b(\tilde{\bm w}[n-1])$. Let ${\bm \Delta}[n]$ be the random vector
denoting which ball is added at round $n$, i.e., a vector of all zeros except for the
index of the urn in which the ball is added, whose entry equals 1.
We have
\begin{eqnarray*}
\bar{\bm w}[n] = \E[\tilde{\bm w}[n]] =
\E\left[\frac{(n_0+n-1) \tilde{\bm w}[n-1] + \bm{\Delta}[n]}{n_0+n}\right] =
\frac{(n_0+n-1)\bar{\bm w}[n-1] + {\bm b}[n]}{n_0+n}.
\end{eqnarray*}
Therefore, from $n-1$ to $n$, $\bar{\bm w}[n]$ moves within domain ${\mathcal S}$ along the
segment connecting $\bar{\bm w}[n-1]$ to ${\bm b}[n]$.
Note that ${\bm b}[n]$ can lie within a different subregion of ${\mathcal S}$ with respect
to vector $\bar{\bm w}[n-1]$. In that case, $\bar{\bm w}[n]$ will eventually enter
the subregion of ${\bm b}[n]$, and since then start moving towards a possible different
attraction point ${\bm b}$. Instead,
if ${\bm b}[n]$ lies in the same subregion of ${\mathcal S}$ as $\bar{\bm w}[n-1]$,
the trajectory will end up at point ${\bm b}[n]$, which must be a fixed point of function
$B(.)$. Figure \ref{fig:exampleN3} illustrates examples of trajectories on the plane $b_3,b_2$
for the system $N=3$, $K=2$, $\alpha=1$, {\em with-item-repetition}. In this case simplex ${\mathcal S}$ is partitioned
into 6 subregions, and there are two stable points ${\bm e}_1 = (\frac{4}{121},\frac{21}{121},\frac{96}{121})$
and ${\bm e}_2 = (\frac{4}{121},\frac{60}{121},\frac{57}{121})$. The associated
permutation graph is shown on the right of Figure \ref{fig:exampleN3}.

\begin{figure}[htb]
\centering
\includegraphics[width=8cm]{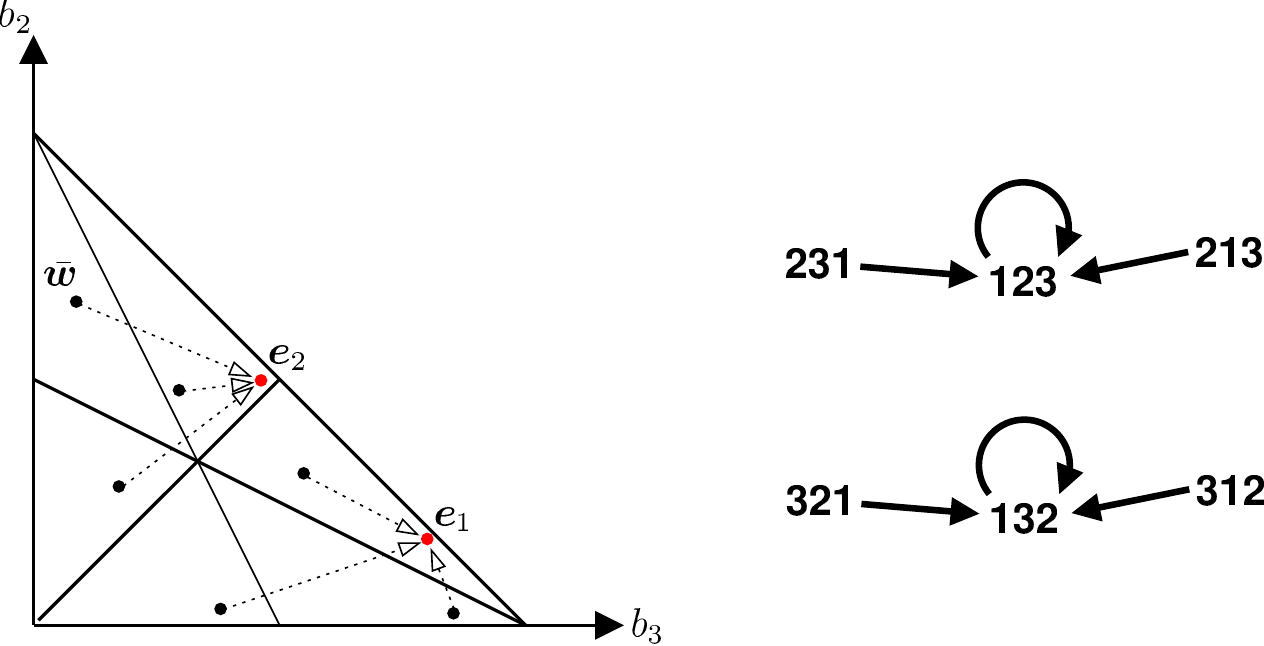}
\caption{Example trajectories  (left) and
permutation graph (right) for the system $N=3$, $K=2$, $\alpha=1$, {\em with-item-repetition}.
The attraction basin of both stable points ${\bm e}_1$ and ${\bm e}_2$ is half of ${\mathcal S}$.\label{fig:exampleN3}}
\end{figure}

Special fixed points of $B(.)$ lying exactly on the hyperspaces in which two or more
components are equal are not stable, due to the assumption that $B(.)$ is an injective function
of the rank of weights ${\tilde{\bm w}}$. Indeed, small perturbations
around such points, which naturally occurs in the actual urn model,
would drive the system towards different attraction points.

To show that fixed points of $B(.)$ are the only possible stable points of the system,
it remains to prove that the system cannot have a limit cycle, or, in other words,
the associated permutation graph is a DAG (directed acyclic graph). This fact can be
proven by contradiction: suppose that the system, as $n \rightarrow \infty$, visits in sequence subregions
${\mathcal S}_1, {\mathcal S}_2,\ldots, {\mathcal S}_C$, $C \geq 2$, characterized by
winning probabilities ${\bm b}_i$, $1 \leq i \leq C$, and staying an asymptotic fraction
$f_i$ of time in subregion $i$, $1 \leq i \leq C$. Then normalized weights
will converge, as $n \rightarrow \infty$, to the unique point
${\bm b}_\infty = \sum_{i=1}^{C} f_i {\bm b}_i$, hence it cannot
keep visiting different attraction points ${\bm b}_i$.
\end{proof}

\section{Proof of Lemma \ref{lemma:critical}}
\label{app:critical}

\begin{proof}
We analyze the stability of a generic permutation in which the $v$-th highest quality item
is the first one (in the natural sequence with increasing quality) that is not given
the right popularity according to the natural permutation.
In the natural permutation, such item (the $v$-th highest quality) would get
popularity $[v^\alpha G]^{-1}$, and suppose instead that it gets
higher popularity $[(v-\Delta)^\alpha G]^{-1}$, with $\Delta \geq 1$,
in the generic alternate permutation. For the alternate permutation to be stable,
the winning probability $b_v$ of this first disordered item must be higher than the winning
probability of all items receiving lower popularity, and most crucially of the item
that gets popularity value $[(v-\Delta+1)^\alpha G]^{-1}$, whose quality
cannot be smaller than that of the $(v-1)$-th highest quality item,
since items with quality rank $v+1,v+2,\ldots,N$ are given their natural popularity,
by hypothesis.
In this situation, the popularities assigned to items with quality rank $N,N-1,\ldots,v,v-1$, are, respectively:
$$ \frac{1}{N^\alpha G}, \frac{1}{(N-1)^\alpha G}, \ldots, \frac{1}{(v+1)^\alpha G}, \frac{1}{(v-\Delta)^\alpha G}, \frac{1}{(v-\Delta+1)^\alpha G}.$$
Let $s^*(v) = \sum_{i = v+1}^{N} (i^\alpha G)^{-1}$, and notice that this is the cumulative
popularity of all items with quality rank higher than $v$, i.e., those items
which receive by hypothesis their correct popularity according to the natural permutation.
The winning probability of the first disordered item is then
$$b_v = \left[ s^*(v) + \frac{1}{(v-\Delta)^\alpha G}\right]^K - [s^*(v)]^K$$
while the winning probability of the item with quality rank $v-1$ is:
\begin{eqnarray*}
b_{v-1} = \left[ s^*(v) + \frac{1}{(v-\Delta)^\alpha G} + \frac{1}{(v-\Delta+1)^\alpha G} \right]^K -
\left[ s^*(v) + \frac{1}{(v-\Delta)^\alpha G}\right]^K.
\end{eqnarray*}
The alternate permutation is stable provided that $b_v > b_{v-1}$, therefore we are left
to equivalently analyze properties of function $F(\Delta,v,K)$:
\begin{equation}\label{eq:Fdvk}
F(\Delta,v,K) = 2 \left[s^*(v) + \frac{1}{(v-\Delta)^\alpha G}\right]^K - [s^*(v)]^K -
\left[ s^*(v) + \frac{1}{(v-\Delta)^\alpha G} + \frac{1}{(v-\Delta+1)^\alpha G} \right]^K
\end{equation}
where integers $\Delta,v,K$ can be chosen in the domain ${\mathcal D}: \Delta \in [1,N-1], v \in [\Delta+1,N], K \geq 1$,
and we are especially interested in possible monotonies of $F(\Delta,v,K)$ in the subdomain ${\mathcal D}^+ \subseteq {\mathcal D}$
in which $F(\Delta,v,K)$ is positive.

We observe that, in the special case $K=1$:
$$ F(\Delta,v,1) = \frac{1}{G}\left[\frac{1}{(v-\Delta)^\alpha} - \frac{1}{(v-\Delta+1)^\alpha}\right] > 0
\qquad \forall (\Delta,v) \in {\mathcal D}$$
which also suggests that, for $K=1$, any local swap of two items, starting from the natural permutation, leads to a stable
alternate permutation. On the other hand, as $K \rightarrow \infty$, $F(\Delta,v,K)$ becomes eventually negative, $\forall (\Delta,v)$,
meaning that, for sufficiently large $K$, any alternate permutation w.r.t. the natural one becomes unstable,
and therefore there exists indeed $K_{\min}$ such that for $K\ \geq K_{\min}$ the natural permutation is the only stable one.
It remains to understand, for fixed $K$, which pair $(\Delta,v)$ provides the largest (positive) $F(\Delta,v,K)$ in ${\mathcal D}^+$.
It turns out that, for fixed $\Delta$, $F(\Delta,v,K)$ is a decreasing function of $v$ in subdomain ${\mathcal D}^+$.
This means that, for given displacement $\Delta$ of the first disordered item from its natural position,
$F(\Delta,v,K)$ achieves its maximum for the minimum $v = \Delta+1$.
Moreover, $F(\Delta,\Delta+1,K)$ is a decreasing function of $\Delta$, and therefore it achieves its maximum for the minimum
$\Delta = 1$. In conclusion, $F(\Delta,v,K)$ achieves its maximum for $\Delta = 1$, $v = \Delta+1$,
hence $K_{\min}$ is determined by the {\em critical} permutation in which the top two highest quality
objects are swapped.
\end{proof}

\section{Proof of Proposition \ref{prop:kmin}}
\label{app:kmin}

\begin{proof}
For $N=1$, we trivially have $K_{\min} = 1$. For $N \geq 2$, properties of $K_{\min}$ can be proven by applying basic calculus
to the real function of two real variables $F(x,G): [1,\infty]\times[1+2^{-\alpha},\infty] \rightarrow \R$:
$$F(x,G) = 2 (1-G^{-1} 2^{-\alpha})^x - 1 - (1-G^{-1}-G^{-1}2^{-\alpha})^x$$
obtained by rewriting inequality \equaref{leftright} relaxing integer $K$ to the real value $x \geq 1$.
We note that $F(x,G)$ is continuous over its domain in both $x$ and $G$.
Moreover $F(1,G)$ is positive for any $\alpha > 0$, while $F(x,G) \rightarrow -1$ as $x \rightarrow \infty$.
Computing $\frac{\partial F}{\partial x}$, one can easily see that there is at most one
value of $x$ at which $\frac{\partial F}{\partial x} = 0$, which is enough
to conclude that, for $\alpha > 0$, there exists a unique value $x^* > 1$ at which \mbox{$F(x^*,G) = 0$}, \mbox{$\frac{\partial F}{\partial x}|_{x=x^*} < 0$},
before which $F(x,G)$ is positive, and after which $F(x,G)$ is negative.
Since $K$ must be an integer, we take the smallest integer larger than or equal to $x^*$, i.e., $K_{\min} = \lceil x^* \rceil$.
In the special case $\alpha = 0$, $F(1,G) = 0$ and $F(x,G)$ is monotonically decreasing in $x$, and we
can use $K_{\min} = 2$ to guarantee that \mbox{$b_{N-1} < b_N$}, for any $N$.

With respect to $G$, we observe that $F(x,G) \rightarrow 0^+$ as $G \rightarrow \infty$, and by
computing $\frac{\partial F}{\partial G}$, we find that there is only one
value of $G$ at which $\frac{\partial F}{\partial G} = 0$. This is enough to conclude
that \mbox{$\frac{\partial F}{\partial G}|_{x=x^*} > 0$}.
The sign of the derivative of the implicit function $x = g(G)$ (implicitly defined by $F(x,G) = 0$) with respect to $G$
can thus be determined by the implicit function theorem, which provides:
$$ \frac{\diff g}{\diff G} = -\frac{\partial F/\partial G}{\partial F/\partial x} $$ at the points
at which $F(x,G) = 0$. Since \mbox{$\frac{\partial F}{\partial G} > 0$}, while \mbox{$\frac{\partial F}{\partial x} < 0$},
we have that $\frac{\diff g}{\diff G}$ is positive, and since $G$ monotonically increases with $N$ we conclude
that $x^*$ is an increasing function of $N$, and thus $K_{\min} =  \lceil x^* \rceil$ is a non-decreasing function of $N$.

Since, when $0 < \alpha \leq 1$, the Dirichlet series $G = \sum_{i=1}^{\infty} i^{-\alpha}$ diverges, it follows
that, for $0 < \alpha \leq 1$, $K_{\min} \rightarrow \infty$ as $N \rightarrow \infty$.
Instead, when $\alpha > 1$, as $N \rightarrow \infty$ $G$ converges to Riemann zeta function $\zeta(\alpha)$.
Therefore, $K_{\min}^\infty \coloneqq  \lceil g(\zeta(\alpha) \rceil)$ is an upper bound
to $K_{\min}$ for any $N$, when $\alpha > 1$.
\end{proof}

\section{Proof of Proposition \ref{prop:aveq}}
\label{app:aveq}
\begin{proof}
We will need the following auxiliary lemma which can be proven by elementary calculus.
\begin{lemma}\label{lemma:monot}
Given any two positive integers $N$ and $K$, with $N>K$, the function $f(\alpha): \R^+ \cup {0} \rightarrow (0,1)$:
$$ f(\alpha) = \frac{\sum_{i=1}^{K} i^{-\alpha}}{\sum_{j=1}^{N} j^{-\alpha}} $$
is monotonically increasing with $\alpha$.
\end{lemma}
\begin{proof}
\begin{eqnarray*}
\frac{\diff f(\alpha)}{\diff \alpha} = \frac{1}{G^2} \left( \sum_{j=1}^{N} j^{-\alpha} \sum_{i=1}^{K} (-\log(i)) i^{-\alpha}
- \sum_{j=1}^{N} (-\log(j)) j^{-\alpha} \sum_{i=1}^{K} i^{-\alpha} \right).
\end{eqnarray*}
The term in parenthesis can be simplified to:
$$ \sum_{i=1}^{K} i^{-\alpha} \sum_{j=K+1}^{N} \log(j) j^{-\alpha} - \sum_{i=1}^{K} \log(i) i^{-\alpha} \sum_{j=K+1}^{N} j^{-\alpha}$$
which is positive since, for any $i$, we take indexes $j > i$, and thus:
$$ \sum_{i=1}^{K} i^{-\alpha} \sum_{j=K+1}^{N} \log(j) j^{-\alpha} > \sum_{i=1}^{K} i^{-\alpha} \sum_{j=K+1}^{N} \log(i) j^{-\alpha}.$$
\end{proof}
Going back to Proposition \ref{prop:aveq}, from the definition of $\overline{q}$:
\begin{equation}\label{eq:Nminus}
\overline{q} = \sum_{i=1}^{N} i\,b_i = \sum_{i=1}^{N} i (s_i^K - s_{i-1}^K) = N - \sum_{i=1}^{N-1} s_i^K
\end{equation}
where $s_i = \sum_{j=1}^i \frac{(N-j+1)^{-\alpha}}{G}$.
Computing the derivative of the above expression with respect to $\alpha$, we have
$$ \frac{\partial \overline{q}}{\partial \alpha} = -K  \sum_{i=1}^{N-1} s_i^{K-1} \frac{\partial s_i}{\partial \alpha} $$
which is positive provided that $\frac{\partial s_i}{\partial \alpha}$ is negative, for any $i$.
Since
$$s_i = 1 - \frac{\sum_{j=1}^{N-i} j^{-\alpha}}{\sum_{j=1}^{N} j^{-\alpha}}$$
we can apply Lemma \ref{lemma:monot} with $K = N-i$ to conclude
that $s_i$ is indeed a decreasing function of $\alpha$, for any $i = 1 \ldots N-1$.

In the special case $\alpha = 0$, we have $s_i = i/N$, and we obtain the
minimum possible value
$$ \overline{q}_{\min} = N - \frac{\sum_{i=1}^{N-1} i^K}{N^K}.$$
As $\alpha \rightarrow \infty$, all cumulants $s_i$, $i = 1 \ldots N-1$ vanish,
and $\overline{q}$ approaches (from below) the limit $N$.

For given $N$ and $\alpha$, the fact that $\overline{q}$ increases with $K$ stems
directly from \equaref{Nminus}, since each term $s_i$ of the summation, being smaller than 1, decreases as we make $K$ larger.
\end{proof}

\end{document}